\documentclass[12pt,a4paper]{article}
\usepackage[centertags]{amsmath}
\usepackage{amsbsy}
\usepackage{amssymb}
\usepackage{graphicx}
\numberwithin{equation}{section}
\begin{document}
\null
\begin{flushright}
\begin{tabular}{r}
DFTT 32/97\\
KIAS-P97004\\
hep-ph/9709439
\end{tabular}
\end{flushright}
\begin{center}
\Large \bfseries
ATMOSPHERIC NEUTRINO OSCILLATIONS 
WITH THREE NEUTRINOS AND A MASS HIERARCHY
\\[1cm]
\large \mdseries
C. Giunti$^{\mathrm{(a)}}$,
C.W. Kim$^{\mathrm{(b)}}$
and
M. Monteno$^{\mathrm{(c)}}$
\\[0.5cm]
\itshape
\large
$^{\mathrm{(a)}}$
\normalsize
INFN, Sezione di Torino, 
and
Dipartimento di Fisica Teorica,
\\
Universit\`a di Torino,
Via P. Giuria 1, I--10125 Torino, Italy
\\[0.3cm]
\large
$^{\mathrm{(b)}}$
\normalsize
Department of Physics and Astronomy,
The Johns Hopkins University,
\\
Baltimore, Maryland 21218, USA, and
\\
School of Physics,
Korea Institute for Advanced Study
\\
Seoul 130-012, Korea
\\[0.3cm]
\large
$^{\mathrm{(c)}}$
\normalsize
INFN, Sezione di Torino, 
and
Dipartimento di Fisica Sperimentale,
\\
Universit\`a di Torino,
Via P. Giuria 1, I--10125 Torino, Italy
\\
\vspace*{1cm}
\upshape
\large
Abstract
\\[0.5cm]
\normalsize
\begin{minipage}[t]{0.9\textwidth}
A comprehensive formalism for the description
of neutrino oscillations
in the Earth
in  a general scheme with three
massive neutrinos
and the mass hierarchy
$ m_1 \ll m_2 \ll m_3 $
is presented.
Using this formalism,
which is valid both in vacuum and in a medium,
the matter effect on the oscillations of
low-energy neutrinos is discussed,
pointing out the existence of
very long oscillations
which are independent of  the
neutrino masses and the neutrino energy,
and are very sensitive to
the matter density along the neutrino trajectory.
As an example of application of  the formulation,
a fit of the Kamiokande atmospheric neutrino data
with the matter effect taken into account
for neutrinos propagating in the Earth
is presented.
The results of the fit indicate that
$
4 \times 10^{-3} \, \mathrm{eV}^2
\lesssim
m_3^2
\lesssim
4 \times 10^{-2} \, \mathrm{eV}^2
$
and the oscillation amplitudes in all channels
($\nu_{\mu}\leftrightarrows\nu_{e}$,
 $\nu_{\mu}\leftrightarrows\nu_{\tau}$,
 $\nu_{e}\leftrightarrows\nu_{\tau}$)
could be large.
Hence,
long-baseline experiments
with reactor
(CHOOZ and Palo Verde)
and accelerator
(K2K, MINOS and ICARUS)
neutrinos
could observe neutrino oscillations
in all channels
with a relatively large statistics.
\end{minipage}
\end{center}

\newpage
\renewcommand{\arraystretch}{1.5}

\section{Introduction}
\label{Introduction}

The atmospheric neutrino anomaly
\cite{KAM88,KAM92,KAM94,IMB,Soudan}
is one of the indications
in favor of the existence of neutrino oscillations
\cite{Pontecorvo57}
(see Refs.\cite{BP78,BP87,Mohapatra-Pal,CWKim,GR95}),
a phenomenon
which is realized
if neutrinos are massive particles
and leptons,
as in the case of  quarks,
are mixed
(i.e. the mass matrices
of the weak-interacting leptons
are not diagonal).
Neutrino oscillation
is considered today
one of the most interesting phenomena
in high energy physics
as well as one of the most promising ways
to search new physics beyond
the Standard Model
(see Ref.\cite{Mohapatra-Pal,GR95}).

The atmospheric neutrino anomaly
has been observed by the
Kamiokande
\cite{KAM88,KAM92,KAM94},
IMB
\cite{IMB}
and Soudan
\cite{Soudan}
experiments,
which measured a ratio
of contained muon-like to electron-like events
smaller than the expected one
\cite{BGS,Gaisser90,Naumov,Honda,Perkins}.
The experimental data can be explained by
$\nu_\mu\to\nu_\tau$
or
$\nu_\mu\leftrightarrows\nu_e$
oscillations with a neutrino mass-squared difference
$\Delta{m}^{2} \sim 10^{-2}\, \mathrm{eV}^2 $.
A preliminary analysis of recent SuperKamiokande data
\cite{SK-atm}
confirms the atmospheric neutrino anomaly,
but indicates a smaller value of $\Delta{m}^{2}$:
$
3 \times 10^{-4}
\lesssim
\Delta m^2
\lesssim
6 \times 10^{-3} \, \mbox{eV}^2
$.
The anomaly observed by Kamiokande and SuperKamiokande
cannot be explained with a background
flux of slow neutrons
\cite{OR,KAM96,SK-atm}.
In this paper,
for the sake of definiteness,
we use only the Kamiokande data,
waiting for definitive data
from SuperKamiokande.

On the other hand,
no anomaly was observed 
in the ratio
of contained muon-like to electron-like events
by the
Fr\'ejus
\cite{Frejus}
and
NUSEX
\cite{NUSEX}
experiments and
in the ratio of contained muon-like to total events
with energy higher than 1 GeV
by the IMB experiment \cite{IMB97}.
However,
the statistical error of the Fr\'ejus and NUSEX data is
higher than that of the Kamiokande data
and the inclusion in the fit of the data
of the Fr\'ejus and NUSEX experiments
does not eliminate the necessity of neutrino oscillations
\cite{Fogli1,BGK}.
Also the high energy IMB data have relatively large errors
and they are not statistically incompatible with the data of the
Kamiokande experiment
\cite{IMB97}.
No anomaly was also observed 
in the absolute flux of upward-going muons by the
Kamiokande
\cite{Kam-atm-up},
IMB
\cite{IMB-up},
Baksan
\cite{BAKSAN}
and
MACRO\footnote{
Although the total flux of upward-going muons
observed by the MACRO experiment is compatible,
within errors,
with the calculated one
(without neutrino oscillations),
the angular shape of the flux does not match the expectations,
because of a deficit in the angular bins
corresponding to the vertical direction.
However,
the compatibility of this deficit with neutrino oscillations
is still uncertain
\cite{MACRO}.
}
\cite{MACRO}
experiments.
However,
the knowledge of the
absolute value of the flux of
neutrino-induced upward-going muons
without neutrino oscillations
is rather model--dependent
\cite{NGS}
(see Ref.\cite{FLM97}
for a combined analysis of the available
upward-going muon data).

The neutrino oscillation solution
of the atmospheric neutrino anomaly
will be checked in the next years
by the
CHOOZ
\cite{CHOOZ}
and
Palo Verde
\cite{PaloVerde}
long-baseline
$\bar\nu_e$
disappearance experiments with reactor anti-neutrinos
and by the
KEK--SuperKamiokande (K2K) \cite{K2K},
Fermilab--Soudan (MINOS) \cite{MINOS} and
CERN--Gran Sasso (ICARUS)
\cite{ICARUS}
long-baseline experiments
with accelerator neutrinos and anti-neutrinos,
which are sensitive to the disappearance of
$
\stackrel{\makebox[0pt][l]
{$\hskip-3pt\scriptscriptstyle(-)$}}{\nu_{\mu}}
$
and to
$
\stackrel{\makebox[0pt][l]
{$\hskip-3pt\scriptscriptstyle(-)$}}{\nu_{\mu}}
\to
\stackrel{\makebox[0pt][l]
{$\hskip-3pt\scriptscriptstyle(-)$}}{\nu_{e}}
$
and
$
\stackrel{\makebox[0pt][l]
{$\hskip-3pt\scriptscriptstyle(-)$}}{\nu_{\mu}}
\to
\stackrel{\makebox[0pt][l]
{$\hskip-3pt\scriptscriptstyle(-)$}}{\nu_{\tau}}
$
transitions.

Another important indication in favor of neutrino oscillations
comes from the results of the solar
neutrino experiments
(Homestake \cite{Homestake},
Kamiokande \cite{Kamiokande},
GALLEX \cite{GALLEX},
SAGE \cite{SAGE}
and
SuperKamiokande \cite{SK-sun}),
which observed event
rates significantly smaller than the values predicted by the
Standard Solar Model (SSM)
\cite{Bahcall,Saclay,CDF,DS96}.
Moreover,
the experimental data
indicate that the suppression of solar $\nu_e$'s
depends on energy,
excluding an astrophysical solution
of the solar neutrino problem
\cite{phenomenological}.
Assuming the validity of the SSM,
the experimental data can be
described by MSW resonant transitions
\cite{MSW}
with
$ \Delta{m}^{2} \sim 10^{-5} \, \mathrm{eV}^2 $
\cite{SOLMSW}
or by vacuum oscillations
with
$ \Delta{m}^{2} \sim 10^{-10} \, \mathrm{eV}^2 $
\cite{SOLVAC}
($\Delta{m}^{2}$
is a neutrino mass-squared difference).

A solution of the solar and atmospheric
neutrino problem with neutrino oscillations
requires the existence of two different scales of
neutrino mass-squared differences,
which corresponds to
the existence of three massive neutrinos,
$\nu_1$,
$\nu_2$
and
$\nu_3$,
with masses
$m_1$,
$m_2$
and
$m_3$,
respectively.
These three massive neutrinos
are mixings of the three flavor neutrinos,
$\nu_e$,
$\nu_\mu$
and
$\nu_\tau$,
whose existence is known\footnote{
The tau neutrino
has not been directly observed,
but it is widely believed to exist without doubt
because it is necessary for the consistency
of the Standard Model
of electro-weak interactions.
}
from the measurements
of the invisible width of the $Z$ boson
done by LEP experiments
(see Ref.\cite{PDG}).

The experimental upper bounds
on the values of the neutrino masses
(see Refs.\cite{Breviews,PDG})
imply that the neutrino masses are much smaller
than the masses of the corresponding charged leptons.
An attractive theoretical explanation
of this smallness
is given by the see-saw mechanism
\cite{see-saw},
which allows the hierarchical pattern
(see Refs.\cite{BP87,Mohapatra-Pal,CWKim,GR95})
\begin{equation}
m_1 \ll m_2 \ll m_3
\,.
\label{pattern}
\end{equation}
This is the simplest and most natural scheme
for the neutrino masses,
 analogous to the mass-schemes of
charged leptons, up and down quarks,
all of which have a hierarchy of masses.
The hierarchical scheme (\ref{pattern})
is exactly what is needed for
a neutrino oscillation solution of
the solar and atmospheric neutrino problems,
if
$
m_2^2
\sim
10^{-5} \, \mbox{or} \, 10^{-10} \, \mathrm{eV}^2
$
(for the MSW or vacuum oscillation solution
of the solar neutrino problem)
and
$
m_3^2
\sim
10^{-2} \, \mathrm{eV}^2
$
(for the solution
of the atmospheric neutrino anomaly).
In this paper we consider this possibility
\cite{Fogli1,BGK}
and we present a comprehensive 
formalism
for the description of
atmospheric neutrino oscillations, by
taking into account
the matter effect
\cite{Wolfenstein78,MSW}
for neutrinos propagating in the Earth
\cite{matter,KP89}.
We believe that this  formalism will be useful
for the analysis of the
atmospheric neutrino data
of the next generation of experiments,
especially those of the SuperKamiokande experiment.

Recently the LSND
collaboration reported
\cite{LSND}
the observation of anomalous
$ \bar\nu_e \, p \to e^+ \, n $
events produced by neutrinos
originating from $\mu^+$ decays
at rest.
These events could be explained by
$ \bar\nu_\mu \to \bar\nu_e $
oscillations with
$ \Delta{m}^{2} \sim 1 \, \mathrm{eV}^2 $,
but this possibility is excluded
in the schemes with three neutrinos
if the neutrino masses
have the values appropriate
for the explanation of the
solar and atmospheric neutrino problem
\cite{no3}.
In this paper we do not consider the LSND
indication in favor of neutrino oscillations,
waiting for its confirmation
by other experiments
(KARMEN \cite{KARMEN}
and others
\cite{LSNDcheck}).

In Section \ref{Analysis}
we present our fit of the Kamiokande data
in the scheme (\ref{pattern}),
with the assumption that
$
m_2^2
\sim
10^{-5} \, \mbox{or} \, 10^{-10} \, \mathrm{eV}^2
$,
so that
$ \Delta{m}^{2}_{21} \equiv m_2^2 - m_1^2 \simeq m_2^2 $
is relevant for the oscillations of solar neutrinos.
With these assumptions,
the oscillations of atmospheric neutrinos
depend only on
$ \Delta{m}^{2}_{31} \equiv m_3^2 - m_1^2 \simeq m_3^2 $.

In the analysis of the experimental indications
in favor of neutrino oscillations
one must take into account also the
numerous negative
results of the terrestrial neutrino oscillation experiments
with reactor
$\bar\nu_{e}$'s
and accelerator
$\nu_\mu$'s and $\bar\nu_\mu$'s
(the latest results are summarized
in Ref.\cite{Boehm-Vannucci}).
In particular,
in Section \ref{Results of the fit}
it will be shown that
part of the region of neutrino mixing parameters
allowed by the atmospheric neutrino data
is excluded by the results of the
Bugey \cite{Bugey}
and
Krasnoyarsk \cite{Krasnoyarsk}
reactor
$\bar\nu_e$
disappearance experiment.

The plan of the paper is as follows.
In Section \ref{Oscillations of 3 neutrinos}
we present the general formalism
for the description of oscillations
of three neutrinos
in vacuum and in matter.
In Section \ref{One mass scale}
we  mainly  discuss  the oscillations
of atmospheric neutrinos
in the scheme (\ref{pattern})
with three neutrinos and a mass hierarchy.
In Section \ref{Low energy neutrinos}
we discuss the matter effect
on the oscillations of neutrinos with low energy.
In Section \ref{Analysis}
we present our fit of
the Kamiokande atmospheric data.
In Section \ref{Conclusions}
we draw our conclusions.
In the Appendix \ref{The mixing matrix}
we present a derivation
of the parameterization of the mixing matrix
used in the paper.

\section{Oscillations of 3 neutrinos}
\label{Oscillations of 3 neutrinos}

The theory of neutrino mixing
(see Refs.\cite{BP78,BP87,Mohapatra-Pal,CWKim,GR95})
is based on the hypothesis that
the left-handed flavor neutrino fields
$\nu_{\alpha L}$
are superpositions
of the left-handed components
of (Dirac or Majorana)
massive neutrino fields
$\nu_{kL}$:
\begin{equation}
\nu_{\alpha L}
=
\sum_{k}
U_{\alpha k}
\nu_{kL}
\,,
\label{000}
\end{equation}
where $U$ is a unitary mixing matrix
($
U U^{\dagger}
=
U^{\dagger} U
=
1
$).

The results of the LEP experiments
on the measurement of the invisible 
width of the $Z$ boson
(see Ref.\cite{PDG})
imply that only three "light"
active flavor neutrinos exist in nature, i.e.
$\nu_{e}$,
$\nu_{\mu}$
and
$\nu_{\tau}$.
However,
the existence
of sterile flavor neutrinos
is not ruled out
and the number of neutrinos
with a definite mass
is unknown.
We will consider here the
simplest scenario
with only
three
active flavor neutrinos
and
three 
(Dirac or Majorana) 
neutrinos with definite mass.
Hence,
in the following
we will consider
the flavor indices
$ \alpha , \beta , \rho , \sigma = e , \mu , \tau $
and the mass indices
$ k , j = 1 , 2 , 3 $.

A neutrino
with momentum $p$
produced by
a charged-current weak interaction process
together with a charged lepton
$ \alpha $
is described
by a flavor state
$
\left|
\nu_{\alpha}(p)
\right\rangle
$,
which is a superposition
of neutrino mass eigenstates
$
\left|
\nu_{k}(p)
\right\rangle
$:
\begin{equation}
\left| \nu_{\alpha}(p) \right\rangle
=
\sum_{k}
U_{{\alpha}k}^{*}
\left| \nu_{k}(p) \right\rangle
\,.
\label{001}
\end{equation}
The state
$ \left| \nu_{k}(p) \right\rangle $
describes a neutrino with a definite mass $m_k$,
momentum $p$,
energy
$ E_{k} = \sqrt{ p^2 + m_{k}^2 } $,
and satisfies the energy eigenvalue equation
\begin{equation}
\mathcal{H}_{0}
\left|
\nu_{k}(p)
\right\rangle
=
E_{k}
\left|
\nu_{k}(p)
\right\rangle
\,,
\label{0011}
\end{equation}
where
$\mathcal{H}_{0}$
is the free neutrino Hamiltonian.
Since
the neutrino masses are very small,
the detectable neutrinos
are extremely relativistic
and
the approximation
$ E_{k} \simeq p + m_{k}^2 / 2 p $
is allowed.

In this paper we consider
atmospheric neutrinos,
which are produced at the top
of the atmosphere by the interaction
of primary cosmic rays with the nuclei in the air
(see Ref.\cite{Gaisser90}).
Atmospheric neutrinos
propagate in the atmosphere,
which is practically equivalent to vacuum,
and in the interior of the Earth,
where the matter density
is sufficiently high
to modify the effective energy-momentum
dispersion relation
through the coherent interaction with
the particles in the medium
\cite{Wolfenstein78,matter,KP89}.
In the following part of this Section we
derive the evolution
equation
for the transition amplitudes
between flavor states
which is valid both in vacuum and in matter.

For neutrinos propagating in matter,
the Hamiltonian $\mathcal{H}$ is given by
\begin{equation}
\mathcal{H}
=
\mathcal{H}_{0}
+
\mathcal{H}_{I}
\,,
\label{0012}
\end{equation}
where $\mathcal{H}_{I}$
is the effective weak interaction Hamiltonian
due to the coherent interaction with the
electrons, protons and neutrons
in  a  medium.
The flavor states
(\ref{001})
are eigenstates of
$\mathcal{H}_{I}$:
\begin{equation}
\mathcal{H}_{I}
\left|
\nu_{\alpha}(p)
\right\rangle
=
V_{\alpha}
\left|
\nu_{\alpha}(p)
\right\rangle
\,,
\label{0013}
\end{equation}
with
\begin{equation}
V_{e}
=
V_{CC} + V_{NC}
\,,
\qquad \qquad
V_{\mu}
=
V_{\tau}
=
V_{NC}
\,.
\label{0014}
\end{equation}
Here
$V_{CC}$ and $V_{NC}$
are,
respectively,
the charged-current
and neutral-current effective potentials
of neutrinos,
given by
\begin{equation}
V_{CC}
=
\sqrt{2}
\,
G_{F}
\,
N_{e}
\,,
\qquad \qquad
V_{NC}
=
-
\frac{ \sqrt{2} }{ 2 }
\,
G_{F}
\,
N_{n}
\,,
\label{00141}
\end{equation}
where
$G_{F}$
is the Fermi constant
and
$N_{e}$
and
$N_{n}$
are,
respectively,
the electron and neutron number densities
of the medium\footnote{
Here we consider an electrically neutral medium,
in which the number density of protons
is equal to the number density of electrons.
This implies that the neutral-current effective potentials
of protons and electrons cancel each other.}.
The effective potentials of anti-neutrinos have
the same absolute value,
but opposite sign:
$\bar{V}_{CC}=-V_{CC}$,
$\bar{V}_{NC}=-V_{NC}$.

Let us consider a
neutrino
with momentum $p$
produced at the time $t=0$ by
a charged-current weak interaction process
together with a charged lepton
$ \alpha $
($ \alpha = e , \mu , \tau $).
In the Schr\"odinger picture,
at the time $t$
this neutrino
is described
by the state
\begin{equation}
\left| \varphi^{(\alpha)}(p,t) \right\rangle
=
\sum_{\beta}
\varphi^{(\alpha)}_{\beta}(p,t)
\left| \nu_{\beta}(p) \right\rangle
\,.
\label{0015}
\end{equation}
This state is a superposition
of flavor states
$\left| \nu_{\beta}(p) \right\rangle$
with amplitudes
$ \varphi^{(\alpha)}_{\beta}(p,t) $
which depend on time
and have the initial value
$ \varphi^{(\alpha)}_{\beta}(p,0) = \delta_{\alpha\beta} $.

The time evolution of the state (\ref{0015})
is given by the Schr\"odinger equation
\begin{equation}
\begin{split}
i
\,
\frac{ \mathrm{d} }{ \mathrm{d}t }
\left| \varphi^{(\alpha)}(p,t) \right\rangle
\null & \null = \null
\mathcal{H}
\left| \varphi^{(\alpha)}(p,t) \right\rangle
=
\sum_{\rho}
\varphi^{(\alpha)}_{\rho}(p,t)
\left(
\mathcal{H}_{0}
+
\mathcal{H}_{I}
\right)
\left| \nu_{\rho}(p) \right\rangle
\\
\null & \null = \null
\sum_{\rho}
\varphi^{(\alpha)}_{\rho}(p,t)
\left(
\sum_{k}
U_{{\rho}k}^{*}
\,
E_{k}
\left| \nu_{k}(p) \right\rangle
+
V_{\rho}
\left| \nu_{\rho}(p) \right\rangle
\right)
\\
\null & \null = \null
\sum_{\sigma,\rho}
\varphi^{(\alpha)}_{\rho}(p,t)
\left(
\sum_{k}
U_{{\rho}k}^{*}
\,
E_{k}
\,
U_{{\sigma}k}
+
V_{\rho}
\,
\delta_{\rho\sigma}
\right)
\left| \nu_{\sigma}(p) \right\rangle
\,.
\end{split}
\label{002}
\end{equation}
Projecting this equation on
$ \left\langle \nu_{\beta}(p) \right| $
and taking into account that
$
\langle
\nu_{\beta}(p)
|
\nu_{\rho}(p)
\rangle
=
\delta_{\beta\rho}
$,
we obtain
\begin{equation}
i \,
\frac{ \mathrm{d} }{ \mathrm{d}t }
\,
\varphi^{(\alpha)}_{\beta}(p,t)
=
\sum_{\rho}
\left(
\sum_{k}
U_{{\beta}k}
\,
E_{k}
\,
U_{{\rho}k}^{*}
+
V_{\rho}
\,
\delta_{\beta\rho}
\right)
\varphi^{(\alpha)}_{\rho}(p,t)
\,.
\label{0021}
\end{equation}
This is the evolution equation for the flavor amplitudes
$\varphi^{(\alpha)}_{\beta}(p,t)$,
whose squared-modulus
gives the probability of
$ \nu_{\alpha} \to \nu_{\beta} $
transitions:
$
P_{\nu_{\alpha}\to\nu_{\beta}}(p,t)
=
\left| \varphi^{(\alpha)}_{\beta}(p,t) \right|^2
$.

For relativistic neutrinos
we have
\begin{equation}
i \,
\frac{ \mathrm{d} }{ \mathrm{d}t }
\,
\varphi^{(\alpha)}_{\beta}(p,t)
=
\left( p + V_{NC} \right)
\varphi^{(\alpha)}_{\beta}(p,t)
+
\sum_{\rho}
\left(
\sum_{k}
U_{{\beta}k}
\,
\frac{ m_{k}^2 }{ 2 \, p }
\,
U_{{\rho}k}^{*}
+
V_{CC}
\,
\delta_{{\beta}e}
\,
\delta_{{\rho}e}
\right)
\varphi^{(\alpha)}_{\rho}(p,t)
\,,
\label{0022}
\end{equation}
where we have separated the contribution
of the
neutral-current effective potential
$V_{NC}$,
which is the same for the three neutrino flavors,
from the contribution
of the
charged-current effective potential
$V_{CC}$,
which affects only the electron neutrino component.
The time $t$
is equal (in natural units) to the
distance $L$ of propagation of the neutrino.

Since neutrino oscillations
are due to the
difference of the phases of different mass eigenstates,
in the following we consider the simplified evolution
equation for the amplitudes
\begin{equation}
\psi^{(\alpha)}_{\beta}(p,t) \equiv
\varphi^{(\alpha)}_{\beta}(p,t)
\,
\exp\!\left( i p t + i \int_{0}^{t} V_{NC}(t') \mathrm{d}t' \right)
\,,
\label{01111}
\end{equation}
where we have taken into account the fact
that $V_{NC}$
may not be  constant along the neutrino trajectory.
For relativistic neutrinos we have
\begin{equation}
i \,
\frac{ \mathrm{d} }{ \mathrm{d}t }
\,
\psi^{(\alpha)}_{\beta}(p,t)
=
\sum_{\rho}
\left(
\sum_{k}
U_{{\beta}k}
\,
\frac{ m_{k}^2 }{ 2 \, p }
\,
U_{{\rho}k}^{*}
+
V_{CC}
\,
\delta_{{\beta}e}
\,
\delta_{{\rho}e}
\right)
\psi^{(\alpha)}_{\rho}(p,t)
\,.
\label{011}
\end{equation}
This is the evolution equation for the transition amplitudes
$\psi^{(\alpha)}_{\beta}(p,t)$,
which allows one to calculate
the oscillation probabilities
in vacuum
($ V_{CC} = 0 $)
as well as in matter
($ V_{CC} \neq 0 $).
The probability to detect a neutrino with flavor $\beta$
after a time $t$,
which corresponds to a propagation distance
$L=t$,
is given by
\begin{equation}
P_{\nu_{\alpha}\to\nu_{\beta}}(p,t)
=
\left| \varphi^{(\alpha)}_{\beta}(p,t) \right|^2
=
\left| \psi^{(\alpha)}_{\beta}(p,t) \right|^2
\,.
\label{012}
\end{equation}

The solution of the evolution equation (\ref{011})
in vacuum is simply given by
\begin{equation}
\psi^{(\alpha)}_{\beta}(p,t)
=
\sum_{k}
U_{{\alpha}k}^{*}
\,
U_{{\beta}k}
\,
\exp\left(
- i
\,
\frac{ m_{k}^2 }{ 2 \, p }
\,
t
\right)
\,.
\label{0111}
\end{equation}
The application of Eq.(\ref{012})
leads to the well known formula for the oscillation probabilities
\begin{equation}
P_{\nu_{\alpha}\to\nu_{\beta}}(p,t)
=
\sum_{k}
|U_{{\alpha}k}|^2
\,
|U_{{\beta}k}|^2
+
2
\,
\mathrm{Re}
\sum_{k>j}
U_{{\alpha}k}^{*}
\,
U_{{\beta}k}
\,
U_{{\alpha}j}
\,
U_{{\beta}j}^{*}
\,
\exp\left(
- i
\,
\frac{ \Delta{m}^2_{kj} }{ 2 \, p }
\,
t
\right)
\,,
\label{0112}
\end{equation}
with
$ \Delta{m}^2_{kj} \equiv m_k^2 - m_j^2 $.

We  now derive  a method for the solution of the
evolution equation (\ref{011})
in matter.
It is convenient to write the evolution equation
(\ref{011}) in
matrix form:
\begin{equation}
i
\,
\frac{ \mathrm{d} }{ \mathrm{d}t }
\,
\Psi^{(\alpha)}_{W}(p,t)
=
\frac{ 1 }{ 2 \, p }
\left(
U
\,
M^2
\,
U^{\dagger}
+
A_{W}
\right)
\Psi^{(\alpha)}_{W}(p,t)
\,,
\label{013}
\end{equation}
where
\begin{equation}
\Psi^{(\alpha)}_{W}(p,t)
\equiv
\begin{pmatrix}
\psi^{(\alpha)}_{e}(p,t)
\\
\psi^{(\alpha)}_{\mu}(p,t)
\\
\psi^{(\alpha)}_{\tau}(p,t)
\end{pmatrix}
\,,
\quad
M
\equiv
\mathrm{diag}( m_{1} , m_{2} , m_{3} )
\,,
\quad
A_{W}
\equiv
\mathrm{diag}( A_{CC} , 0 , 0 )
\,,
\label{014}
\end{equation}
and
$ A_{CC} \equiv 2 p V_{CC} $.
For anti-neutrinos,
$ A_{CC} $
must be replaced by
$ \bar{A}_{CC} = - A_{CC} $.

The $3\times3$ mixing matrix $U$
can be written in the form
(see the Appendix \ref{The mixing matrix})
\begin{equation}
U
=
V_{23}
\,
V_{13}
\,
W_{12}
\,
D(\lambda)
\,,
\label{015}
\end{equation}
with the orthogonal matrices
\begin{equation}
V_{23}
=
\begin{pmatrix}
1 & 0 & 0
\\
0 & \cos\vartheta_{23} & \sin\vartheta_{23}
\\
0 & -\sin\vartheta_{23} & \cos\vartheta_{23}
\end{pmatrix}
\,,
\qquad
V_{13}
=
\begin{pmatrix}
\cos\vartheta_{13} & 0 & \sin\vartheta_{13}
\\
0 & 1 & 0
\\
-\sin\vartheta_{13} & 0 & \cos\vartheta_{13}
\end{pmatrix}
\,,
\label{0161}
\end{equation}
and the unitary matrices
\begin{equation}
W_{12}
=
D_{12}
\,
V_{12}
\,
D_{12}^{\dagger}
\,,
\quad
D(\lambda)
=
\mathrm{diag}\!\left(
e^{i\lambda_{1}}
\, , \,
e^{i\lambda_{2}}
\, , \,
1
\right)
\,,
\label{0162}
\end{equation}
with
\begin{equation}
V_{12}
=
\begin{pmatrix}
\cos\vartheta_{12} & \sin\vartheta_{12} & 0
\\
-\sin\vartheta_{12} & \cos\vartheta_{12} & 0
\\
0 & 0 & 1
\end{pmatrix}
\,,
\quad
D_{12}
=
\mathrm{diag}\!\left(
e^{i\eta_{12}}
\, , \,
1
\, , \,
1
\right)
\,.
\label{0164}
\end{equation}
Here
$\vartheta_{23}$,
$\vartheta_{13}$,
and
$\vartheta_{12}$
are the three mixing angles,
$\eta_{12}$
is the Dirac CP-violating phase
\cite{KM73}
and
$\lambda_{1}$
and
$\lambda_{2}$
are the two
Majorana CP-violating phases.
The matrix
$V_{ab}$
represents a rotation
of an angle
$ \vartheta_{ab} $
in the
$ \nu_{a} $--$ \nu_{b} $
plane.
In the case of Dirac neutrinos
the matrix
$D(\lambda)$
containing the two Majorana CP-violating phases
can be eliminated with a suitable redefinition
of the arbitrary phases
of the Dirac neutrino fields.
This operation is not possible for Majorana neutrinos
because the Majorana mass term
is not invariant under rephasing
of the neutrino fields.
However,
the presence of the matrix
$D(\lambda)$
does not have any effect
on neutrino oscillations
in vacuum
\cite{BHP80,Doi81}
as well as in matter
\cite{Langacker87}.
Indeed,
from Eq.(\ref{013})
it is clear that
neutrino oscillations depend on the quantity
$
U
\,
M^2
\,
U^{\dagger}
$
and,
since the diagonal matrix
$M^2$
commutes
with the diagonal matrix
$D(\lambda)$,
we have
\begin{equation}
U
\,
M^2
\,
U^{\dagger}
=
V_{23}
\,
V_{13}
\,
W_{12}
\,
D(\lambda)
\,
M^2
\,
D(\lambda)^{\dagger}
\,
W_{12}^{\dagger}
\,
V_{13}^{\dagger}
\,
V_{23}^{\dagger}
=
V_{23}
\,
V_{13}
\,
W_{12}
\,
M^2
\,
W_{12}^{\dagger}
\,
V_{13}^{\dagger}
\,
V_{23}^{\dagger}
\,.
\label{0163}
\end{equation}
Contrary to the Majorana phases,
the Dirac phase
$\eta_{12}$,
contained in
$W_{12}$,
has an effect on neutrino oscillations
\cite{Cabibbo78}.
Under CP transformations
$ U \xrightarrow{\mathrm{CP}} U^{*} $,
which is equivalent to
$ \eta_{12} \xrightarrow{\mathrm{CP}} -\eta_{12} $.
This means that
CP violation could be observed
in neutrino oscillations,
measuring the differences
$
P_{\nu_{\alpha}\to\nu_{\beta}}
-
P_{\bar\nu_{\alpha}\to\bar\nu_{\beta}}
$
for
$\alpha\neq\beta$
(CPT invariance
implies that
$
P_{\nu_{\alpha}\to\nu_{\beta}}
=
P_{\bar\nu_{\beta}\to\bar\nu_{\alpha}}
$
and therefore
$
P_{\nu_{\alpha}\to\nu_{\alpha}}
-
P_{\bar\nu_{\alpha}\to\bar\nu_{\alpha}}
$
is always equal to zero).
Assuming CPT invariance,
a violation of CP implies a violation of T,
which could be observed
by measuring the difference
$
P_{\nu_{\alpha}\to\nu_{\beta}}
=
P_{\nu_{\beta}\to\nu_{\alpha}}
$
for
$\alpha\neq\beta$.

The fact that the matrix 
$ A_{W} $
has only one non-zero element
$ (A_{W})_{11} = A_{CC} $
implies that
\begin{equation}
V_{23}^{\dagger}
\,
A_{W}
\,
V_{23}
=
A_{W}
\,.
\label{20104}
\end{equation}
Therefore,
it is convenient to define the new
column matrix of amplitudes
\begin{equation}
\widetilde{\Psi}^{(\alpha)}(p,t)
\equiv
V_{23}^{\dagger}
\,
\Psi^{(\alpha)}_{W}(p,t)
\,,
\label{20105}
\end{equation}
which satisfy the simplified evolution equation
\begin{equation}
i
\,
\frac{ \mathrm{d} }{ \mathrm{d}t }
\,
\widetilde{\Psi}^{(\alpha)}(p,t)
=
\frac{ \widetilde{M}^{2} }{ 2 \, p }
\,
\widetilde{\Psi}^{(\alpha)}(p,t)
\,,
\label{20106}
\end{equation}
with
\begin{equation}
\widetilde{M}^{2}
=
V_{13}
\,
W_{12}
\,
M^2
\,
W_{12}^{\dagger}
\,
V_{13}^{\dagger}
+
A_{W}
\,.
\label{20107}
\end{equation}
Notice that the amplitudes in the column matrix
$\widetilde{\Psi}^{(\alpha)}(p,t)$
do not have a definite flavor or mass character.
They are introduced only as a tool for the solution
of the evolution equation (\ref{013}).

The standard procedure that has been 
used  in finding the solution
of Eq.(\ref{20106})
consists  of  the diagonalization
of the effective squared-mass matrix
$ \widetilde{M}^{2} $
(see Refs.\cite{KP89,CWKim}).
This method is appropriate
for a constant matter density
or for a medium
whose density
changes along the path of neutrino propagation
are much slower
than the changes of the oscillation phases
(adiabatic approximation).
Techniques have been developed in order
to extend the solution to
non-adiabatic cases
in which the matter density have a smooth and monotonic
variation
(as is the case for solar neutrinos;
see Refs.\cite{KP89,CWKim}).
However,
the general case of irregular
matter variations
must be solved by integrating numerically
Eq.(\ref{20106})
(or directly Eq.(\ref{013})).

In the following we consider neutrino oscillations
in the Earth,
whose internal composition is well approximated
by a number of shells with constant density
(see Refs.\cite{Stacey,Anderson}).
The solid curve in Fig.\ref{fig1}A
shows the density $\rho$
in the interior of the Earth
as a function of the radial distance $r$
from the center of the Earth
according to the data given in Ref.\cite{Anderson}.
The dotted curve represents
our approximation in terms of five shells
with constant densities
$
\rho_{i=1,\ldots,5}
=
13.0,11.3,5.0,3.9,3.0
\, \mathrm{g}/\mathrm{cm}^3
$,
with outer radii
$
r_{i=1,\ldots,5}
=
1221,3480,5701,5971,6371
\, \mathrm{Km}
$.
The solid curve in Fig.\ref{fig1}B
shows the electron number density
$ N_e = \rho \langle Z / A \rangle $
as a function of the radial distance $r$
and
the dotted curve represents
our approximation in terms of five shells
with constant electron number density
$
(N_e)_{i=1,\ldots,5}
=
6.15,5.36,2.47,1.93,1.50
\, N_A \, \mathrm{cm}^{-3}
$.
For the averaged ratio
$ \langle Z / A \rangle $
we took
$ \langle Z / A \rangle_{1,2} = 0.475 $
for the two inner shells (core)
and
$ \langle Z / A \rangle_{3,4,5} = 0.495 $
for the three outer shells (mantle).

The effective squared-mass matrix
$ \widetilde{M}^{2} $
can be diagonalized separately in each shell,
yielding a relatively simple solution
for the evolution equation
(\ref{20106})
in the shell under consideration.
The solutions in confining shells
are matched on the shell boundaries
by the continuity of the flavor states,
which imply the continuity
of the amplitudes
$\Psi^{(\alpha)}_{W}(p,t)$
and,
from Eq.(\ref{20105}),
of the amplitudes
$\widetilde{\Psi}^{(\alpha)}(p,t)$
(the transformation (\ref{20105})
does not depend on the matter density).

In the following
we consider a scheme with
the hierarchy (\ref{pattern})
for the neutrino masses.
In this case the oscillations
of atmospheric neutrinos depend 
only  on one mass scale.
This scheme allows a relatively simple
diagonalization
of the $ 3 \times 3 $
effective squared-mass matrix
$ \widetilde{M}^{2} $.

\subsection{One mass scale}
\label{One mass scale}

We will consider here a scheme
with three neutrinos and
the hierarchy (\ref{pattern})
for the neutrino masses.
We will assume that
the squared-mass difference
$
\Delta{m}^2_{21}
\equiv
m_{2}^2 - m_{1}^2
\simeq
m_2^2
$
is relevant for the explanation of
the solar neutrino problem
($ \Delta{m}^2_{21} \sim 10^{-5} \, \mathrm{eV}^2 $
\cite{SOLMSW}
for MSW resonant transitions
or
$ \Delta{m}^2_{21} \sim 10^{-10} \, \mathrm{eV}^2 $
\cite{SOLVAC}
for vacuum oscillations).
In this case
\begin{equation}
\frac{ m_1^2 \, R_{\oplus} }{ 2 \, p }
\ll
1
\,,
\qquad
\frac{ m_2^2 \, R_{\oplus} }{ 2 \, p }
\ll
1
\,,
\label{2020111}
\end{equation}
where
$ R_{\oplus} = 6371 \, \mathrm{Km} $
is the radius of the Earth,
which represents
a characteristic distance
of propagation for atmospheric neutrinos.
Hence,
the phases generated by
$ m_{1}^2 $
and
$ m_{2}^2 $
can be neglected for atmospheric neutrinos
and the squared-mass matrix $M^2$
can be approximated with
\begin{equation}
M^2
=
\mathrm{diag}(0,0,m_{3}^2)
\,.
\label{202011}
\end{equation}
Before proceeding with the derivation of the
oscillation probabilities that follow from this approximation,
it is necessary to notice that
caution is needed for low-energy atmospheric neutrinos if
$ m_2^2 \gtrsim 10^{-5} \, \mathrm{eV}^2 $,
because in this case
$ m_2^2 R_{\oplus} / 2 p \ll 1 $
only for
$ p \gg 150 \, \mathrm{MeV} $.
According to the most recent analyses of the solar neutrino data
\cite{recent-sun},
including preliminary data from SuperKamiokande
\cite{SK-sun},
the small and large mixing angle MSW solutions
of the solar neutrino problem
require,
respectively,
$
4 \times 10^{-6} \, \mathrm{eV}^2
\lesssim
m_2^2
\lesssim
1.2 \times 10^{-5} \, \mathrm{eV}^2
$
and
$
9 \times 10^{-6} \, \mathrm{eV}^2
\lesssim
m_2^2
\lesssim
3 \times 10^{-5} \, \mathrm{eV}^2
$
at 90\% CL.
Hence,
expecially in the case of the large mixing angle MSW solution
of the solar neutrino problem,
the formalism described in this section may be not
applicable to low-energy atmospheric neutrinos
(as, for example, some part of those contributing to the
Kamiokande and SuperKamiokande sub-GeV data).
Since,
as will be discussed later,
the formalism described in this section
is very convenient for the analysis
of atmospheric neutrino data in the case of
three-neutrino mixing
(the oscillation probabilities are independent from
$\vartheta_{12}$ and $\eta_{12}$),
we think that if a high value of $m_2^2$
will be established,
it will be convenient to analyze the atmospheric neutrino data
with a cut in energy such that
$ m_2^2 R_{\oplus} / 2 p \ll 1 $ 
(at least for a first analysis,
before the calculation of a complete fit
which should include also
the solar neutrino data
for the determination of
$m_2^2$ and $\vartheta_{12}$).
Let us emphasize that there are good hopes that the value of
$m_2^2 \simeq \Delta{m}^2_{21}$
will be determined by the new generation of solar neutrino experiments
(SuperKamiokande,
SNO,
ICARUS,
Borexino,
GNO
and others
\cite{future-sun})
that are expected to be able to distinguish
among the different solutions of the solar neutrino problem.

The simple form
of the squared-mass matrix (\ref{202011})
allows one to write the
oscillation probabilities in vacuum in an elegant way:
using the unitarity relation
\begin{equation}
\sum_{k=1,2}
U_{{\beta}k}
U_{{\alpha}k}^{*}
=
\delta_{\alpha\beta}
-
U_{{\beta}3}
U_{{\alpha}3}^{*}
\,,
\label{22021}
\end{equation}
the amplitudes (\ref{0111})
can be written as
\begin{equation}
\begin{split}
\psi^{(\alpha)}_{\beta}(p,t)
\null & \null = \null
\left(
\sum_{k=1,2}
U_{{\alpha}k}^{*}
\,
U_{{\beta}k}
\right)
+
U_{{\alpha}3}^{*}
\,
U_{{\beta}3}
\,
\exp\left(
- i
\,
\frac{ m_{3}^2 \, t }{ 2 \, p }
\right)
\\
\null & \null = \null
\delta_{\alpha\beta}
+
U_{{\alpha}3}^{*}
\,
U_{{\beta}3}
\left[
\exp\left(
- i
\,
\frac{ m_{3}^2 \, t }{ 2 \, p }
\right)
-
1
\right]
\,.
\end{split}
\label{2202}
\end{equation}
For the oscillation probabilities
we obtain
\cite{BFP92,BBGK}
\begin{subequations}
\label{2107}
\begin{align}
\null & \null
P_{\nu_{\alpha}\to\nu_{\beta}}
=
A_{{\alpha};{\beta}}
\,
\sin^2\!\left(
\frac{ m_{3}^2 \, t }{ 4 \, p }
\right)
\qquad \qquad
(\beta\neq\alpha)
\,,
\label{21071}
\\
\null & \null
P_{\nu_{\alpha}\to\nu_{\alpha}}
=
1
-
\sum_{\beta\neq\alpha}
P_{\nu_{\alpha}\to\nu_{\beta}}
=
1
-
B_{{\alpha};{\alpha}}
\,
\sin^2\!\left(
\frac{ m_{3}^2 \, t }{ 4 \, p }
\right)
\,,
\label{21072}
\end{align}
\end{subequations}
where
\begin{subequations}
\label{2108}
\begin{align}
\null & \null
A_{{\alpha};{\beta}}
=
4
\,
|U_{{\alpha}3}|^2
\,
|U_{{\beta}3}|^2
\,,
\label{21081}
\\
\null & \null
B_{{\alpha};{\alpha}}
=
\sum_{\beta\not=\alpha}
A_{{\alpha};{\beta}}
=
4
\left| U_{\alpha3} \right|^2
\left(
1
-
\left| U_{\alpha3} \right|^2
\right)
\label{21082}
\end{align}
\end{subequations}
are the
classical oscillation amplitudes.

It is important to notice the following features of
the oscillation probabilities (\ref{2107}):

\begin{enumerate}

\item
All oscillation channels
($\nu_{\mu}\leftrightarrows\nu_{e}$,
 $\nu_{\mu}\leftrightarrows\nu_{\tau}$,
 $\nu_{e}\leftrightarrows\nu_{\tau}$)
are open
and have
the same oscillation length
\begin{equation}
L_{\mathrm{osc}}
=
\frac{ 4 \pi p }{ m_3^2 }
\,.
\label{280291}
\end{equation}

\item
The transition probabilities
are determined by three parameters:
$ m_3^2 $,
$ |U_{e3}|^2 $
and
$ \left| U_{\mu3} \right|^2 $
(from the unitarity of the mixing matrix it follows that
$
|U_{\tau3}|^2
=
1
-
|U_{e3}|^2
-
|U_{\mu3}|^2
$).
The expression of
$ |U_{e3}|^2 $
and
$ \left| U_{\mu3} \right|^2 $
in terms of the mixing angles
of the parameterization (\ref{015})
is
\begin{equation}
|U_{e3}|^2
=
\sin^2 \vartheta_{13}
\,,
\qquad
|U_{\mu3}|^2
=
\cos^2 \vartheta_{13}
\,
\sin^2 \vartheta_{23}
\,.
\label{2802}
\end{equation}
Hence,
the oscillation probabilities
depend on the two mixing angles
$\vartheta_{13}$
and
$\vartheta_{23}$
and
do not depend on the value of the mixing angle
$\vartheta_{12}$
and
of the Dirac CP violating phase
$\eta_{12}$.
Since this fact is due to the degeneracy
of the first two squared mass eigenvalues
in Eq.(\ref{202011}),
it remains valid for the oscillations in matter
(see Eq.(\ref{202121})).

\item
The expressions (\ref{2107})
have the same form
as the standard expressions
for the oscillation probabilities
in the case of mixing between
two massive neutrino fields
(see Refs.\cite{BP78,BP87,Mohapatra-Pal,CWKim,GR95}):
\begin{subequations}
\label{21077}
\begin{align}
\null & \null
P_{\nu_{\alpha}\to\nu_{\beta}}
=
\sin^2 2\vartheta
\,
\sin^2\!\left(
\frac{ \Delta{m}^2 \, t }{ 4 \, p }
\right)
\qquad \qquad
(\beta\neq\alpha)
\,,
\label{210771}
\\
\null & \null
P_{\nu_{\alpha}\to\nu_{\alpha}}
=
1
-
\sin^2 2\vartheta
\,
\sin^2\!\left(
\frac{ \Delta{m}^2 \, t }{ 4 \, p }
\right)
\,,
\label{210772}
\end{align}
\end{subequations}
where
$\Delta{m}^2$
is the neutrino mass-squared difference
and
$\vartheta$
is the two-generation mixing angle.
In the case of two-neutrino mixing
only oscillations between two flavors are possible
and
the oscillation probabilities
are characterized by two parameters,
$ \Delta m^2 $
(which determines the oscillation length
$
L_{\mathrm{osc}}
=
4 \pi p / \Delta m^2
$)
and
$ \sin^2 2\vartheta $.
The important difference between
the two-neutrino mixing scheme
and the scheme with
three-neutrino mixing
and a mass hierarchy that we consider here
is that
the second scheme
allows simultaneous transitions
among all three flavor neutrinos
($\nu_{\mu}\leftrightarrows\nu_{e}$,
 $\nu_{\mu}\leftrightarrows\nu_{\tau}$,
 $\nu_{e}\leftrightarrows\nu_{\tau}$).

\item
The equality in  the form of the
oscillation probabilities
in the two-neutrino mixing scheme
and in the three-neutrino mixing scheme
(\ref{pattern})
is very important,
because the data of all terrestrial
oscillation experiments
with reactor and accelerator neutrinos
have been analyzed by the experimental collaborations
under the assumption
of two-generation mixing,
obtaining constraints on the possible values of
the mixing parameters
$\Delta{m}^{2}$
and
$\sin^{2}2\vartheta$.
The results of the analysis of the experimental data
are presented in the form of allowed (or excluded)
regions
in the
$\sin^{2}2\vartheta$--$\Delta{m}^{2}$
plane.
Identifying the appropriate
$A_{\alpha;\beta}$
or
$B_{\alpha;\alpha}$
with
$\sin^{2}2\vartheta$
and
$m_3^2$
with
$\Delta{m}^{2}$,
the results of the standard analyses of
the neutrino oscillation data
yield allowed (or excluded)
regions in the
$A_{\alpha;\beta}$--$m_3^2$
and
$B_{\alpha;\alpha}$--$m_3^2$
planes
(with $\alpha,\beta=e,\mu,\tau$).

\item
Since
the classical oscillation amplitudes (\ref{2108})
depend only on the squared moduli
of the elements of the mixing matrix
and
$
A_{{\beta};{\alpha}}
=
A_{{\alpha};{\beta}}
$,
it is clear that
\begin{subequations}
\label{2801}
\begin{align}
\null & \null
P_{\nu_{\alpha}\to\nu_{\beta}}
=
P_{\bar\nu_{\alpha}\to\bar\nu_{\beta}}
\,,
\label{28011}
\\
\null & \null
P_{\nu_{\alpha}\to\nu_{\beta}}
=
P_{\nu_{\beta}\to\nu_{\alpha}}
\,.
\label{28012}
\end{align}
\end{subequations}
Hence,
if neutrino oscillations depend  only  on one mass scale,
CP and T violations are not observable.
This is true for neutrino oscillations
in vacuum as well as in matter,
although in matter Eq.(\ref{28011})
is no longer satisfied
(see Eq.(\ref{20205})
and the following discussion).

\end{enumerate}

Let us now consider
oscillations in matter
and
discuss how to solve
the evolution equation (\ref{20106}).
The form (\ref{202011})
for the mass matrix implies that
$
W_{12}
M^2
W_{12}^{\dagger}
=
M^2
$
and
the evolution equation (\ref{013})
does not depend on the value of the mixing angle
$\vartheta_{12}$
and of the Dirac CP violating phase
$\eta_{12}$.
Then,
from Eq.(\ref{20107}) we have
\begin{equation}
\widetilde{M}^{2}
=
V_{13}
\,
M^2
\,
V_{13}^{\dagger}
+
A_{W}
=
\left(
{\setlength{\arraycolsep}{5pt}
\begin{array}{ccc} \displaystyle
m_{3}^2 \, \sin^2\vartheta_{13} + A_{CC} &
0 & m_{3}^2 \, \cos\vartheta_{13} \, \sin\vartheta_{13}
\\ \displaystyle
0 & 0 & 0
\\ \displaystyle
m_{3}^2 \, \cos\vartheta_{13} \, \sin\vartheta_{13} &
0 & m_{3}^2 \, \cos^{2}\vartheta_{13}
\end{array}}
\right)
\label{20205}
\end{equation}
The disappearance of the Dirac CP violating phase
$\eta_{12}$
from the evolution equation
for the transition amplitudes
implies that,
in the scheme under consideration,
CP violation is not observable in neutrino oscillations
in matter, as well as in vacuum.
Let us emphasize,
however,
that,
contrary to the vacuum case,
in matter the survival and transition probabilities
of neutrinos and anti-neutrinos can be different,
because the effective potentials
of neutrinos and anti-neutrinos
have the same absolute value but opposite signs.
Hence,
Eq.(\ref{28011}) is not satisfied in matter.
This is due to the fact that the medium
is not CP invariant and not CPT invariant.
However,
if the matter distribution is symmetric
along the neutrino path,
the matter effect
is T invariant
and
$
P_{\nu_{\alpha}\to\nu_{\beta}}
=
P_{\nu_{\beta}\to\nu_{\alpha}}
$
in matter as in vacuum
(see Eq.(\ref{28012})).

Let us now proceed  to diagonalize 
the matrix
$\widetilde{M}^{2}$
in Eq.(\ref{20205}).
Obviously,
the matrix
$\widetilde{M}^{2}$
has a zero eigenvalue,
$
m^{2}_{M2} = 0
$.
The other two eigenvalues
are given by
\begin{equation}
m^{2}_{M1,3}
=
\frac{ 1 }{ 2 }
\left(
m_{3}^2 + A_{CC}
\right)
\mp
\frac{ 1 }{ 2 }
\sqrt{
\left(
m_{3}^2 \cos2\vartheta_{13} - A_{CC}
\right)^2
+
\left(
m_{3}^2 \sin2\vartheta_{13}
\right)^2
}
\,.
\label{20207}
\end{equation}
These are the effective squared-masses
of neutrinos propagating
in matter.
Notice that the
values of the effective squared-masses
of neutrinos and anti-neutrinos,
which we will denote as
$\bar{m}^{2}_{M1,3}$,
are different,
because
$\bar{A}_{CC}=-A_{CC}$.

Let us define
the column matrix of amplitudes
\begin{equation}
\begin{pmatrix}
\psi^{(\alpha)}_{M1}(p,t)
\\
\psi^{(\alpha)}_{M2}(p,t)
\\
\psi^{(\alpha)}_{M3}(p,t)
\end{pmatrix}
\equiv
\Psi^{(\alpha)}_{M}(p,t)
\equiv
V_{13}^{M\dagger}
\,
\widetilde{\Psi}^{(\alpha)}(p,t)
\,,
\label{20208}
\end{equation}
where the orthogonal matrix
\begin{equation}
V_{13}^{M}
\equiv
\left(
{\setlength{\arraycolsep}{5pt}
\begin{array}{ccc} \displaystyle
\cos\vartheta_{13}^{M} & 0 & \sin\vartheta_{13}^{M}
\\ \displaystyle
0 & 1 & 0
\\ \displaystyle
- \sin\vartheta_{13}^{M} & 0 & \cos\vartheta_{13}^{M}
\end{array}}
\right)
\,,
\label{20209}
\end{equation}
is defined in such a way
that the matrix
\begin{equation}
V_{13}^{M\dagger}
\,
\widetilde{M}^{2}
\,
V_{13}^{M}
\equiv
M^{2}_{M}
\equiv
\mathrm{diag}( m^{2}_{M1} , 0 , m^{2}_{M3} )
\label{20210}
\end{equation}
is diagonal.
Then,
we obtain
that the effective mixing angle in matter
$\vartheta_{13}^{M}$
is given by
\begin{equation}
\cos 2\vartheta_{13}^{M}
=
\frac
{ m_{3}^2 \, \cos2\vartheta_{13} - A_{CC} }
{ \sqrt{
\left( m_{3}^2 \, \cos2\vartheta_{13} - A_{CC} \right)^2
+
\left( m_{3}^2 \, \sin2\vartheta_{13} \right)^2
} }
\,.
\label{20211}
\end{equation}
If
$ \cos2\vartheta_{13} > 0 $,
for
$
A_{CC}
\gg
m_{3}^2
$
the effective mixing angle
$\vartheta_{13}^{M}$
for neutrinos
is approximately equal to $\pi/2$.
The corresponding
effective mixing angle
for anti-neutrinos,
which we will denote as
$\bar\vartheta_{13}^{M}$,
tends to vanish
because $\bar{A}_{CC}$
is negative.
When
$ A_{CC} = m_{3}^2 \, \cos2\vartheta_{13} $
there is a resonance:
$\vartheta_{13}^{M}=\pi/4$
and
the mixing in the 1-3 sector is maximal.
If
$ \cos2\vartheta_{13} < 0 $,
the resonance condition can be realized for anti-neutrinos
($\bar\vartheta_{13}^{M}=\pi/4$)
at
$ \bar{A}_{CC} = \cos2\vartheta_{13} $.
In this case
$\bar\vartheta_{13}^{M}\simeq\pi/2$
and
$\vartheta_{13}^{M}\simeq0$
for
$
A_{CC}
\gg
m_{3}^2
$.
However,
the analogy with the quark sector,
in which there is a hierarchy of mixing
that  respects the mass hierarchy,
 suggests that a small value of
$\vartheta_{13}$
is natural and
that the resonance
condition can be realized for neutrinos.

The evolution equation for
$ \Psi^{(\alpha)}_{M}(p,t) $
is diagonal:
multiplying Eq.(\ref{20106}) by $V_{13}^{M\dagger}$
on the left , 
we have
\begin{equation}
i
\,
\frac{ \mathrm{d} }{ \mathrm{d}t }
\,
\Psi^{(\alpha)}_{M}(p,t)
=
\frac{ M^{2}_{M} }{ 2 \, p }
\,
\Psi^{(\alpha)}_{M}(p,t)
\,,
\label{20213}
\end{equation}
with
$
M^{2}_{M}
=
\mathrm{diag}( m^{2}_{M1} , 0 , m^{2}_{M3} )$.
It is important to notice
that the assumption of a constant matter density
plays here a crucial role.
Indeed,
if the matter density were not constant,
the effective mixing matrix
$V_{13}^{M}$
would be dependent on $t$
and the time derivative
of
$V_{13}^{M}$
would induce additional non-diagonal terms
in the evolution equation (\ref{20213}).

The explicit evolution equation for the amplitudes
$ \psi^{(\alpha)}_{Mk}(p,t) $,
with
$ k=1,2,3 $,
is
\begin{equation}
i
\,
\frac{ \mathrm{d} }{ \mathrm{d}t }
\,
\psi^{(\alpha)}_{Mk}(p,t)
=
\frac{ m^{2}_{Mk} }{ 2 \, p }
\,
\psi^{(\alpha)}_{Mk}(p,t)
\,.
\label{202151}
\end{equation}
This equation has the straightforward solution
\begin{equation}
\psi^{(\alpha)}_{Mk}(p,t)
=
\exp \left[
- i
\,
\frac{ m^{2}_{Mk} }{ 2 \, p }
\left( t - t_{0} \right)
\right]
\psi^{(\alpha)}_{Mk}(p,t_{0})
\,,
\label{202152}
\end{equation}
which can be written in  a matrix form as
\begin{equation}
\Psi^{(\alpha)}_{M}(p,t)
=
\exp \left[
- i
\,
\frac{ M^{2}_{M} }{ 2 \, p }
\left( t - t_{0} \right)
\right]
\Psi^{(\alpha)}_{M}(p,t_{0})
\,.
\label{20215}
\end{equation}
This equation gives the evolution in time of the amplitudes
$\psi^{(\alpha)}_{Mk}(p,t)$.
However,
we are interested in the evolution in time of the amplitudes
$\psi^{(\alpha)}_{\beta}(p,t)$
whose absolute value squared give the
oscillation probabilities
trough Eq.(\ref{012}).
The amplitudes
$\psi^{(\alpha)}_{\beta}(p,t)$
are the elements of
$\Psi^{(\alpha)}_{W}(p,t)$
and,
using Eqs.(\ref{20105}) and (\ref{20208})
we can express
$\Psi^{(\alpha)}_{W}(p,t)$
in terms of
$\Psi^{(\alpha)}_{M}(p,t)$
as
\begin{equation}
\Psi^{(\alpha)}_{W}(p,t)
=
V_{23}
\,
\widetilde{\Psi}^{(\alpha)}(p,t)
=
V_{23}
\,
V_{13}^{M}
\,
\Psi^{(\alpha)}_{M}(p,t)
=
V^{M}
\,
\Psi^{(\alpha)}_{M}(p,t)
\,,
\label{20212}
\end{equation}
with the orthogonal matrix
\begin{equation}
V^{M}
\equiv
V_{23}
\,
V_{13}^{M}
=
\begin{pmatrix}
\cos\vartheta_{13}^{M} & 0 & \sin\vartheta_{13}^{M}
\\
- \sin\vartheta_{23} \sin\vartheta_{13}^{M}
&
\cos\vartheta_{23}
&
\sin\vartheta_{23} \cos\vartheta_{13}^{M}
\\
- \cos\vartheta_{23} \sin\vartheta_{13}^{M}
&
- \sin\vartheta_{23}
&
\cos\vartheta_{23} \cos\vartheta_{13}^{M}
\end{pmatrix}
\,.
\label{202121}
\end{equation}
Notice that the elements of the third column of
$V^{M}$
have the same structure as
the elements of the third column of the mixing matrix
$U$
(see Eq.(\ref{100011})),
in terms
of the mixing angle
$\vartheta_{23}$
and
the effective mixing angle in matter
$\vartheta_{13}^{M}$,
which replaces the vacuum mixing angle
$\vartheta_{13}$.
Let us emphasize,
however,
that the matrix
$V^{M}$
can be considered
as the effective mixing matrix in matter
only for the study of the oscillations in matter
in the scheme
with three neutrinos and the mass hierarchy (\ref{pattern}).
In this case,
the approximation
$m_1=m_2=0$
implies that there is no mixing in the 1-2 sector
and $W_{12}$
is equal to the identity matrix.
Notice that the first row of
$V^{M}$,
corresponding to the electron neutrino,
is particularly simple,
with
$V^{M}_{e2}=0$,
because
the
electron neutrino feels only the mixing in the 1-3 sector.
Physically,
the special status of the electron neutrino
follows from its direct charged current interaction with
the electrons in the medium,
represented by the quantity $A_{CC}$.

Substituting the solution (\ref{20215})
for
$\Psi^{(\alpha)}_{M}(p,t)$
and using the orthogonality of
$V^{M}$,
we obtain
\begin{equation}
\Psi^{(\alpha)}_{W}(p,t)
=
V^{M}
\,
\exp \left[
- i
\,
\frac{ M^{2}_{M} }{ 2 \, p }
\left( t - t_{0} \right)
\right]
(V^{M})^{T}
\,
\Psi^{(\alpha)}_{W}(p,t_{0})
\,.
\label{20216}
\end{equation}
Explicitly, we have
\begin{equation}
\psi^{(\alpha)}_{\beta}(p,t)
=
\sum_{k,\rho}
V^{M}_{{\beta}k}
\,
\exp \left[
- i
\,
\frac{ m^{2}_{Mk} }{ 2 \, p }
\left( t - t_{0} \right)
\right]
V^{M}_{{\rho}k}
\,
\psi^{(\alpha)}_{\rho}(p,t_{0})
\,.
\label{20219}
\end{equation}
This is the solution of the evolution equation
for the amplitudes
$\psi^{(\alpha)}_{\beta}(p,t)$
in a slab of matter with a constant density.
In the case of neutrino propagation
through a series of slabs of matter with a constant density,
as in the interior of the Earth,
the solutions
(\ref{20219})
for confining slabs must be matched
in order to have continuity
of the flavor amplitudes
$\psi^{(\alpha)}_{\beta}(p,t)$.
For example,
for a neutrino created
at the time $t_{0}$
and crossing a series of boundaries
between slabs of matter with a constant density
at the times
$t_{1}$,
$t_{2}$,
\ldots,
$t_{n}$,
the amplitudes
$\psi^{(\alpha)}_{\beta}(p,t)$
in the $(n+1)^{\mathrm{th}}$ slab
are given by the matrix equation
\begin{equation}
\begin{split}
\Psi^{(\alpha)}_{W}(p,t)
\null = \null & \null
\left[
V^{M}
e^{
- i
\frac
{ M^{2}_{M} ( t - t_{n} ) }
{ 2 p }
}
(V^{M})^{T}
\right]_{(n+1)}
\left[
V^{M}
e^{
- i
\frac
{ M^{2}_{M} ( t_{n} - t_{n-1} ) }
{ 2 p }
}
(V^{M})^{T}
\right]_{(n)}
\cdots
\\
\null & \null
\cdots
\left[
V^{M}
e^{
- i
\frac
{ M^{2}_{M} ( t_{2} - t_{1} ) }
{ 2 p }
}
(V^{M})^{T}
\right]_{(2)}
\left[
V^{M}
e^{
- i
\frac
{ M^{2}_{M} ( t_{1} - t_{0} ) }
{ 2 p }
}
(V^{M})^{T}
\right]_{(1)}
\Psi^{(\alpha)}_{W}(p,t_{0})
\,.
\end{split}
\label{202161}
\end{equation}
Here the notation
$[\ldots]_{(i)}$
indicates that the
matter-dependent quantities inside of the square
brackets must be evaluated with the matter density
of the $i^{\mathrm{th}}$ slab.
From Eq.(\ref{012}),
one can see that
the oscillation probabilities
are given by the squared moduli of
the amplitudes
$\psi^{(\alpha)}_{\beta}(p,t)$
given by Eq.(\ref{202161}).

It is clear that for $n>1$
an analytical calculation of
the oscillation probabilities is a rather complicated task and
of little interest.
Therefore,
this calculation
is done numerically with a computer.

The calculation for $n=1$,
i.e.
for the case
of a medium with constant density,
can be done  analytically.
In order to derive the oscillation probabilities,
we employ the same method as in vacuum
(see Eqs.(\ref{22021})--(\ref{21082})):
using the orthogonality relation
\begin{equation}
\sum_{k=1,2}
V^{M}_{{\beta}k}
V^{M}_{{\alpha}k}
=
\delta_{\alpha\beta}
-
V^{M}_{{\beta}3}
V^{M}_{{\alpha}3}
\,,
\label{2202191}
\end{equation}
the amplitudes (\ref{20219}),
with
$t_{0}=0$
and
$ \psi^{(\alpha)}_{\rho}(p,0) = \delta_{\alpha\rho} $,
can be written as
\begin{equation}
\begin{split}
\psi^{(\alpha)}_{\beta}(p,t)
\null = \null & \null
V^{M}_{{\alpha}1}
\,
V^{M}_{{\beta}1}
\,
\exp\left(
- i
\,
\frac{ m_{M1}^2 \, t }{ 2 \, p }
\right)
+
V^{M}_{{\alpha}2}
\,
V^{M}_{{\beta}2}
+
V^{M}_{{\alpha}3}
\,
V^{M}_{{\beta}3}
\,
\exp\left(
- i
\,
\frac{ m_{M3}^2 \, t }{ 2 \, p }
\right)
\\
\null = \null & \null
\exp\left(
- i
\,
\frac{ m_{M1}^2 \, t }{ 2 \, p }
\right)
\left\{
\delta_{\alpha\beta}
+
V^{M}_{{\alpha}2}
\,
V^{M}_{{\beta}2}
\left[
\exp\left(
i
\,
\frac{ m_{M1}^2 \, t }{ 2 \, p }
\right)
-
1
\right]
\right.
\\
\null & \null
\hskip4cm
\left.
+
V^{M}_{{\alpha}3}
\,
V^{M}_{{\beta}3}
\left[
\exp\left(
- i
\,
\frac{ \Delta{m}^{2}_{M31} \, t }{ 2 \, p }
\right)
-
1
\right]
\right\}
\,,
\end{split}
\label{220292}
\end{equation}
with
\begin{equation}
\Delta{m}^{2}_{M31}
\equiv
m^{2}_{M3} - m^{2}_{M1}
=
\sqrt{
\left(
m_{3}^2 \cos2\vartheta_{13} - A_{CC}
\right)^2
+
\left(
m_{3}^2 \sin2\vartheta_{13}
\right)^2
}
\,.
\label{202162512}
\end{equation}
For the transition and survival
probabilities
we obtain
\begin{subequations}
\label{210793}
\begin{align}
\null & \null
P_{\nu_{\alpha}\to\nu_{\beta}}
=
A^{M}_{{\alpha};{\beta}}
\,
S^{2}_{31}
-
4
\,
V^{M}_{{\alpha}1}
\,
V^{M}_{{\beta}1}
\,
V^{M}_{{\alpha}2}
\,
V^{M}_{{\beta}2}
\,
S^{2}_{1}
-
4
\,
V^{M}_{{\alpha}2}
\,
V^{M}_{{\beta}2}
\,
V^{M}_{{\alpha}3}
\,
V^{M}_{{\beta}3}
\left( S^{2}_{3} - S^{2}_{31} \right)
\,,
\label{2107931}
\\
\null & \null
P_{\nu_{\alpha}\to\nu_{\alpha}}
=
1
-
B^{M}_{{\alpha};{\alpha}}
\,
S^{2}_{31}
-
4
\,
(V^{M}_{{\alpha}1})^2
\,
(V^{M}_{{\alpha}2})^2
\,
S^{2}_{1}
-
4
\,
(V^{M}_{{\alpha}2})^2
\,
(V^{M}_{{\alpha}3})^2
\left( S^{2}_{3} - S^{2}_{31} \right)
\,,
\label{2107932}
\end{align}
\end{subequations}
respectively.
Here
\begin{equation}
S^{2}_{1}
\equiv
\sin^{2}\!\left(
\frac{ m^{2}_{M1} t }{ 4 \, p }
\right)
\,,
\qquad
S^{2}_{3}
\equiv
\sin^{2}\!\left(
\frac{ m^{2}_{M3} t }{ 4 \, p }
\right)
\,,
\qquad
S^{2}_{31}
\equiv
\sin^{2}\!\left(
\frac{ \Delta{m}^{2}_{M31} t }{ 4 \, p }
\right)
\,,
\label{202162511}
\end{equation}
and
\begin{subequations}
\label{210894}
\begin{align}
\null & \null
A^{M}_{{\alpha};{\beta}}
=
4
\,
(V^{M}_{{\alpha}3})^2
\,
(V^{M}_{{\beta}3})^2
\,,
\label{2108941}
\\
\null & \null
B^{M}_{{\alpha};{\alpha}}
=
\sum_{\beta\not=\alpha}
A^{M}_{{\alpha};{\beta}}
=
4
(V^{M}_{\alpha3})^2
\left[
1
-
(V^{M}_{\alpha3})^2
\right]
\label{2108942}
\end{align}
\end{subequations}
are the
classical oscillation amplitudes in matter
analogous to the corresponding ones in vacuum,
$A_{{\alpha};{\beta}}$
and
$B_{{\alpha};{\alpha}}$
(see Eqs.(\ref{2108})).
The classical oscillation amplitudes in matter
$A^{M}_{{\alpha};{\beta}}$
and
$B^{M}_{{\alpha};{\alpha}}$
have the same structure as
the classical oscillation amplitudes in vacuum
$A_{{\alpha};{\beta}}$
and
$B_{{\alpha};{\alpha}}$,
in terms
of the mixing angle
$\vartheta_{23}$
and
the effective mixing angle in matter
$\vartheta_{13}^{M}$,
which replaces the vacuum mixing angle
$\vartheta_{13}$.
Therefore,
it is easy to see that
in the limit $V_{CC}\to0$
only the terms involving
$  A^{M}_{{\alpha};{\beta}} \to A_{{\alpha};{\beta}} $
and
$ B^{M}_{{\alpha};{\alpha}} \to B_{{\alpha};{\alpha}} $
survive in Eqs.(\ref{210793})
and the oscillation probabilities
reduce to the oscillation probabilities
in vacuum (\ref{2107}).

It is interesting to notice the following features
of the oscillation probabilities (\ref{210793}):

\begin{enumerate}

\item
The oscillation probabilities
involving the electron neutrino
in the initial and/or final state
are particularly simple,
because
$V^{M}_{e2}=0$:
\begin{equation}
P_{\nu_e\to\nu_e}
=
1
-
B^{M}_{e;e}
\,
S^{2}_{31}
\,,
\qquad
P_{\nu_e\to\nu_\mu}
=
A^{M}_{e;\mu}
\,
S^{2}_{31}
\,,
\label{2021621}
\end{equation}
with
\begin{equation}
B^{M}_{e;e}
=
\sin^2 2\vartheta_{13}^{M}
\,,
\qquad
A^{M}_{e;\mu}
=
\sin^2\vartheta_{23}
\,
\sin^2 2\vartheta_{13}^{M}
\,.
\label{2021622}
\end{equation}
These oscillation probabilities depend
 only on one oscillation length,
\begin{equation}
L_{\mathrm{osc}}^{(31)}
=
\frac{ 4 \pi p }{ \Delta{m}^{2}_{M31} }
\,.
\label{280292}
\end{equation}
In the limit
$V_{CC}\to0$
we have
$ L_{\mathrm{osc}}^{(31)} \to L_{\mathrm{osc}} $,
where
$L_{\mathrm{osc}}$
is the oscillation length in vacuum,
given in Eq.(\ref{280291}).

\item
The oscillation probabilities involving only
muon and/or tau neutrinos have a complicated structure,
depending on three different oscillation lengths:
$L_{\mathrm{osc}}^{(31)}$,
$L_{\mathrm{osc}}^{(3)}$
and
$L_{\mathrm{osc}}^{(1)}$,
with
\begin{equation}
L_{\mathrm{osc}}^{(3)}
=
\frac{ 4 \pi p }{ m^{2}_{M3} }
\,,
\qquad
L_{\mathrm{osc}}^{(1)}
=
\frac{ 4 \pi p }{ m^{2}_{M1} }
\,.
\label{280293}
\end{equation}
In the limit $V_{CC}\to0$,
we have that
$ L_{\mathrm{osc}}^{(31)} \to L_{\mathrm{osc}} $,
$ L_{\mathrm{osc}}^{(3)} \to L_{\mathrm{osc}} $
and
$ L_{\mathrm{osc}}^{(1)} \to \infty $.

The survival probability of muon neutrinos is given by
\begin{equation}
P_{\nu_\mu\to\nu_\mu}
=
1
-
B^{M}_{\mu;\mu}
\,
S^{2}_{31}
-
\sin^2 2\vartheta_{23}
\,
\sin^2\vartheta_{13}^{M}
\,
S^{2}_{1}
-
\sin^2 2\vartheta_{23}
\,
\cos^{2}\vartheta_{13}^{M}
\left( S^{2}_{3} - S^{2}_{31} \right)
\,,
\label{2021623}
\end{equation}
with
\begin{equation}
B^{M}_{\mu;\mu}
=
\sin^4 \vartheta_{23}
\,
\sin^2 2\vartheta_{13}^{M}
+
\sin^2 2\vartheta_{23}
\,
\cos^{2}\vartheta_{13}^{M}
\,.
\label{202164}
\end{equation}

\item
The oscillation probabilities (\ref{210793})
are invariant under time reversal:
$
P_{\nu_{\alpha}\to\nu_{\beta}}
=
P_{\nu_{\beta}\to\nu_{\alpha}}
$.
This is due to the fact that
the effect of a constant matter density
along the neutrino path
is T invariant.
As a consequence,
all the oscillation probabilities
can be derived
from the three probabilities
(\ref{2021621})  and  (\ref{2021623})
using the conservation of probabilities
($
\sum_{\beta}
P_{\nu_{\alpha}\to\nu_{\beta}}
=
1
$)
and the invariance under time reversal
($
P_{\nu_{\alpha}\to\nu_{\beta}}
=
P_{\nu_{\beta}\to\nu_{\alpha}}
$).

\item
The survival and transition probabilities
of neutrinos and anti-neutrinos can be different,
because
the effective mixing angles
$\vartheta_{13}^{M}$,
$\bar\vartheta_{13}^{M}$
and the effective squared-masses
$m^{2}_{M1,3}$,
$\bar{m}^{2}_{M1,3}$
for neutrinos and anti-neutrinos
are different
($\bar{A}_{CC}=-A_{CC}$).
This is a consequence of the fact that the medium
is  neither CP invariant  nor CPT invariant.

\end{enumerate}

When
$ A_{CC} \gtrsim m_{3}^2 $,
which can be realized
with high matter density
and neutrino energy,
the effective mixing angle
$\vartheta_{13}^{M}$
of neutrinos
becomes large and
approaches
$\pi/2$
for
$ A_{CC} \gg m_{3}^2 $
(see Eq.(\ref{20211});
here we assume
$ \cos2\vartheta_{13} > 0 $).
This means that
the matter effect is relevant and can
 dramatically modify the oscillation probabilities.
From Eqs.(\ref{2021621}) and (\ref{2021622})
we see that,
in the case of a medium with constant density,
when
$ A_{CC} \gg m_{3}^2 $
and
$\vartheta_{13}^{M}\simeq\pi/2$
the transition probabilities
of electron neutrinos are suppressed.
The same suppression is realized
for anti-neutrinos,
which have
$\bar\vartheta_{13}^{M}\simeq0$.
Assuming a small value for
$\vartheta_{13}$,
the maximum difference between the neutrino and anti-neutrino
oscillation probabilities
is obtained when the resonance condition
$ A_{CC} = m_{3}^2 \, \cos2\vartheta_{13} $
is satisfied for neutrinos,
which implies that
$\vartheta_{13}^{M}\simeq\pi/4$
for neutrinos,
whereas
$\bar\vartheta_{13}^{M}<\vartheta_{13}$
for anti-neutrinos.
At the resonance the
oscillations of the survival probability of electron neutrinos
are maximal,
$
P_{\nu_e\to\nu_e}
=
1
-
S^{2}_{31}
$,
while the other probabilities depend on the value of
$\vartheta_{23}$.

\subsection{Low energy neutrinos}
\label{Low energy neutrinos}

It has been pointed out
\cite{Pantaleone94,Pantaleone}
that the matter effect may be not negligible even
in the case of low matter density
or low neutrino energy\footnote{
Let us remind that,
as discussed after Eq.(\ref{202011}),
caution is needed for low-energy neutrinos if
$ m_2^2 \gtrsim 10^{-5} \, \mathrm{eV}^2 $.
In this case the considerations presented in this subsection
can be applied only to neutrinos in the momentum range
$ m_2^2 R_{\oplus} / 2 \ll p \ll m_3^2 / 2 V_{CC} $.
For example,
for
$ m_2^2 = 1.5 \times 10^{-5} \, \mathrm{eV}^2 $
(that corresponds to the best-fit point
of the large mixing angle MSW solution
of the solar neutrino problem \cite{recent-sun}),
$ m_3^2 = 10^{-2} \, \mathrm{eV}^2 $
and
$ V_{CC} = 5 \times 10^{-13} \, \mathrm{eV} $
(that corresponds to the density in the core
of the earth)
we have the momentum range
$ 200 \, \mathrm{MeV} \ll p \ll 10 \, \mathrm{GeV} $.
},
i.e.
for $ A_{CC} = 2 p V_{CC} \ll m_{3}^2 $.
In this case,
the mass eigenvalues
can be approximated by
\begin{subequations}
\label{2030}
\begin{align}
&
m^{2}_{M1}
\simeq
A_{CC}
\left( 1 + \cos2\vartheta_{13} \right)
/ 2
\,,
\label{20301}
\\
&
m^{2}_{M3}
\simeq
m_{3}^2
+
A_{CC}
\left( 1 - \cos2\vartheta_{13} \right)
/ 2
\,,
\label{20302}
\\
&
\Delta{m}^{2}_{M31}
\simeq
m_{3}^2
-
A_{CC}
\,
\cos2\vartheta_{13}
\,.
\label{20303}
\end{align}
\end{subequations}
It is important to notice
\cite{Pantaleone}
that
$m^{2}_{M1}$
is proportional to $A_{CC}$,
which  in turn is proportional to the neutrino energy
($ A_{CC} = 2 p V_{CC} $).
Hence,
the phase
\begin{equation}
\frac{ m^{2}_{M1} \, t }{ 2 \, p }
\simeq
\frac{ A_{CC} \, t }{ 4 \, p }
\left( 1 + \cos2\vartheta_{13} \right)
=
\frac{ V_{CC} \, t }{ 2 }
\left( 1 + \cos2\vartheta_{13} \right)
\label{20304}
\end{equation}
is independent  of  the neutrino momentum
and can be relevant for low-energy atmospheric neutrinos
(as those corresponding to the Kamiokande sub-GeV data).
The value of the potential $V_{CC}$ is given by
\begin{equation}
\begin{split}
V_{CC}
=
\sqrt{2} \, G_{F} \, N_e
\null & \null = \null
7.63 \times 10^{-14}
\left(
\frac{ N_e }{ N_A \mathrm{cm}^{-3} }
\right)
\mathrm{eV}
\\
\null & \null = \null
3.87 \times 10^{-4}
\left(
\frac{ N_e }{ N_A \mathrm{cm}^{-3} }
\right)
\mathrm{Km}^{-1}
\,,
\end{split}
\label{20306}
\end{equation}
where
$N_A$
is the Avogadro number.
The interior of the Earth has an electron number density
$N_e$ 
that goes from about
$ 2 \, N_A \, \mathrm{cm}^{-3} $
in the mantle
to about
$ 6 \, N_A \, \mathrm{cm}^{-3} $
in the inner core,
with an average value of about
$ 3 \, N_A \, \mathrm{cm}^{-3} $
(see Fig.\ref{fig1}B).
For a propagation of
$ 10^4 \, \mathrm{Km} $
we have
$ V_{CC} t / 2 \simeq 2 \pi $,
which shows that
the phase (\ref{20304})
could be relevant
for low-energy atmospheric neutrinos.

When
$ A_{CC} \ll m_{3}^2 $,
the oscillation probabilities
depend on two oscillation lengths:
\begin{equation}
L_{\mathrm{osc}}^{(31)}
\simeq
L_{\mathrm{osc}}^{(3)}
\simeq
\frac{ 4 \, \pi \, p }{ m_{3}^2 }
\equiv
L_{\mathrm{osc}}^{\mathrm{short}}
\,,
\qquad
L_{\mathrm{osc}}^{(1)}
\simeq
\frac{ 4 \, \pi }{ V_{CC} \left( 1 + \cos2\vartheta_{13} \right) }
\equiv
L_{\mathrm{osc}}^{\mathrm{long}}
\,.
\label{203061}
\end{equation}
The short oscillation length
$L_{\mathrm{osc}}^{\mathrm{short}}$
coincides
with the oscillation length in vacuum,
whereas the long oscillation length
$L_{\mathrm{osc}}^{\mathrm{long}}$
is due to the matter effect.
For example,
for
$ m_{3}^2 = 10^{-2} \, \mathrm{eV}^2 $
(which is close to the best fit of the atmospheric neutrino data),
$ p = 500 \, \mathrm{MeV} $
(which is in the range of the Kamiokande sub-GeV data),
$ N_{e} = 3 \, N_A \, \mathrm{cm}^{-3} $,
we have
$
L_{\mathrm{osc}}^{\mathrm{short}}
\simeq
10^2 \, \mathrm{Km}
$
and
$
L_{\mathrm{osc}}^{\mathrm{long}}
\simeq
10^4 \, \mathrm{Km}
$.
The oscillating terms
$s_{3}^2$
and
$s_{31}^2$
in Eq.(\ref{202162511})
depend on the short oscillation length
$L_{\mathrm{osc}}^{\mathrm{short}}$
and oscillate very fast on a distance scale
bigger than
$ 10^3 \, \mathrm{Km} $.
Therefore,
one can study the slow oscillations of
the probabilities
due to
$L_{\mathrm{osc}}^{\mathrm{long}}$,
averaging
the probabilities
over the fast oscillations
due to
$L_{\mathrm{osc}}^{\mathrm{short}}$:
\begin{subequations}
\label{202163}
\begin{align}
\left\langle P_{\nu_e\to\nu_e} \right\rangle
\null = \null & \null
1
-
\frac{ 1 }{ 2 }
\,
B^{M}_{e;e}
\,,
\qquad
\left\langle P_{\nu_e\to\nu_\mu} \right\rangle
=
\frac{ 1 }{ 2 }
\,
A^{M}_{e;\mu}
\,,
\label{2021632}
\\
\left\langle P_{\nu_\mu\to\nu_\mu} \right\rangle
\null = \null & \null
1
-
\frac{ 1 }{ 2 }
B^{M}_{\mu;\mu}
-
\sin^2 2\vartheta_{23}
\,
\sin^2\vartheta_{13}^{M}
\,
\sin^2\!\left(
\frac{ t }{ L_{\mathrm{osc}}^{\mathrm{long}} }
\right)
\,,
\label{2021633}
\end{align}
\end{subequations}
with
$B^{M}_{e;e}$,
$A^{M}_{e;\mu}$
and
$B^{M}_{\mu;\mu}$
given by Eqs.(\ref{2021622}) and (\ref{202164}).

One can see that
the averaged value of the survival and transition
probabilities of electron neutrinos
are constant and
very close to the corresponding
averaged probabilities in vacuum
(for
$ A_{CC} \ll m_{3}^2 $
we have
$ \vartheta_{13}^{M} \simeq \vartheta_{13} $,
$ B^{M}_{e;e} \simeq B_{e;e} $
and
$ A^{M}_{e;\mu} \simeq A_{e;\mu} $),
\begin{equation}
\left\langle P_{\nu_e\to\nu_e} \right\rangle_{\mathrm{vac}}
=
1
-
\frac{ 1 }{ 2 }
\,
B_{e;e}
\,,
\qquad
\left\langle P_{\nu_e\to\nu_\mu} \right\rangle_{\mathrm{vac}}
=
\frac{ 1 }{ 2 }
\,
A_{e;\mu}
\,.
\label{20216321}
\end{equation}
On the other hand,
the averaged value of the survival
probability of muon neutrinos
has the additional term
depending on
$L_{\mathrm{osc}}^{\mathrm{long}}$
with respect to the corresponding
averaged probability in vacuum
\begin{equation}
\left\langle P_{\nu_\mu\to\nu_\mu} \right\rangle_{\mathrm{vac}}
=
1
-
\frac{ 1 }{ 2 }
B_{\mu;\mu}
\,.
\label{20216331}
\end{equation}
The behaviour of
$
\left\langle
P_{\nu_\mu\to\nu_\mu}
\right\rangle
$
as a function of the propagation distance $L=t$
in a medium with a constant electron density
$ N_{e} = 3 \, N_A \, \mathrm{cm}^{-3} $
is depicted in
Fig.\ref{fig2}
for neutrinos and anti-neutrinos
and two sets of values of the mixing parameters:
(A)
$|U_{e3}|^2=|U_{\mu3}|^2=1/3$
(which corresponds to maximal mixing
of the three neutrinos)
and
(B)
$|U_{e3}|^2=0.3$,
$|U_{\mu3}|^2=0.5$
(which is close to the best fit
of the Kamiokande data;
see Eq.(\ref{best-fit})).
The solid lines represent
$
\left\langle
P_{\nu_\mu\to\nu_\mu}
\right\rangle
$,
the dashed lines represent
$
\left\langle
P_{\bar\nu_\mu\to\bar\nu_\mu}
\right\rangle
$
and
the dotted lines
represent
the averaged survival probability in vacuum
(\ref{20216331}).
From the comparison with the
averaged survival probability in vacuum,
it is clear that the oscillations in matter
can be rather different than in vacuum
even if
$ A_{CC} \ll m_{3}^2 $.
One can also see a small difference between
the averaged survival probabilities
of muon neutrinos and anti-neutrinos,
due to the fact that
$\vartheta_{13}^{M}$
and
$\bar\vartheta_{13}^{M}$
are not exactly equal to
$\vartheta_{13}$
and different between each other.

In
Fig.\ref{fig3}A,B
we present
the behaviour of
$
\left\langle
P_{\nu_\mu\to\nu_\mu}
\right\rangle
$
and
$
\left\langle
P_{\bar\nu_\mu\to\bar\nu_\mu}
\right\rangle
$
for neutrinos propagating
in the interior of the Earth
as a function of the zenithal angle $\theta$.
These figures
correspond to the
same two sets of values of the mixing parameters
as in Fig.\ref{fig2}.
The values of
$
\left\langle
P_{\nu_\mu\to\nu_\mu}
\right\rangle
$
(and similarly those of
$
\left\langle
P_{\bar\nu_\mu\to\bar\nu_\mu}
\right\rangle
$)
are obtained
assuming
$ m_{3}^2 = 10^{-2} \, \mathrm{eV}^2 $,
$ p = 500 \, \mathrm{MeV} $
and averaging numerically
the probability
$
P_{\nu_\mu\to\nu_\mu}
=
\left| \psi^{(\mu)}_{\mu}(p,t) \right|^2
$,
with
$ \psi^{(\mu)}_{\mu}(p,t) $
given by Eq.(\ref{202161}),
over the fast oscillations corresponding to
$
L_{\mathrm{osc}}^{\mathrm{short}}
\simeq
10^2 \, \mathrm{Km}
$.
Figure \ref{fig3}C
shows the slant depth
as a function of the zenithal angle $\theta$.
One can see that
the irregularities of
$
\left\langle
P_{\nu_\mu\to\nu_\mu}
\right\rangle
$
and
$
\left\langle
P_{\bar\nu_\mu\to\bar\nu_\mu}
\right\rangle
$
correspond to irregularities of the slant depth,
which occur for trajectories
that
graze the boundaries between the different shells
with approximately constant density.

The effect of
$L_{\mathrm{osc}}^{\mathrm{long}}$
will be negligible in the long-baseline experiments
of the next generation
(CHOOZ \cite{CHOOZ},
Palo Verde \cite{PaloVerde},
KEK--SuperKamiokande \cite{K2K},
Fermilab--Soudan \cite{MINOS},
CERN--Gran Sasso \cite{ICARUS})
because the baseline will be shorter than
$ 10^3 \, \mathrm{Km} $.
However,
it is interesting to notice that
$L_{\mathrm{osc}}^{\mathrm{long}}$
is independent from the value of $m_3$
and is very sensitive to the matter density.
Therefore,
the observation of
the effect of
$L_{\mathrm{osc}}^{\mathrm{long}}$
in future very-long-baseline
neutrino oscillation experiments
could be helpful for a study of 
 a detailed tomography of the interior
of the Earth \cite{tomography}.

\section{Analysis of the Kamiokande
atmospheric neutrino data}
\label{Analysis}

In this Section
we present our fit of the Kamiokande
atmospheric neutrino data
in 
 the scheme
with mixing of three neutrinos
and a mass hierarchy
considered in the previous Section.
In this scheme
the probability
$
P_{\nu_{\alpha}\to\nu_{\beta}}
(E,\cos\theta;m_{3}^2,|U_{e3}|^{2},|U_{\mu3}|^{2})
$
of
$\nu_{\alpha}\to\nu_{\beta}$
oscillations
for a neutrino with energy
$ E \simeq p $
arriving at the detector
from a direction
with a zenith angle $\theta$
depends on the three parameters
$m_{3}^2$, $|U_{e3}|^{2}$, $|U_{\mu3}|^{2}$
and on the matter density
along the neutrino trajectory.
(In this article, we do not consider
preliminary data from the SuperKamiokande experiment,
waiting for  more refined data to be available in the future.)

\subsection{The experimental data}

We have analyzed the neutrino-induced
$e$-like and $\mu$-like event rates 
measured by the
Kamiokande \cite{KAM88,KAM92,KAM94}
experiment.

The Kamiokande data sample is divided into two classes:
the low energy sub-GeV data,
including only fully contained events,
and the high energy multi-GeV data,
including both fully and partially contained events. 
Additional information on these events is provided
by their angular distributions, 
divided in five zenith-angle bins. 
In our analysis we fit
the angular distribution
of the
$e$-like and $\mu$-like multi-GeV data
and the total number of
$e$-like and $\mu$-like sub-GeV data
reported in Ref.\cite{KAM94}.
For the sub-GeV data we do not fit
the angular distribution
because there is a poor correlation
between the directions of the neutrino
and the produced charged lepton
\cite{KAM94}.

The Kamiokande collaboration measured
the absolute rate of $e$-like and $\mu$-like
events.
In order to analyze the muon-electron
flavor composition of atmospheric
neutrino events
they presented also
the ratio of ratios
$R_{\mu/e} = (\mu/e)_{\mathrm{exp}}/(\mu/e)_{\mathrm{th}}$,
where
$(\mu/e)_{\mathrm{exp}}$
is the ratio of
$\mu$-like and $e$-like events
measured experimentally
and
$(\mu/e)_{\mathrm{th}}$
is the same ratio calculated theoretically,
without neutrino oscillations.
The Kamiokande results
\begin{align}
R_{\mu/e}^{\mathrm{sub-GeV}}
\null & \null = \null
0.60^{+0.06}_{-0.05} \pm 0.05
\,,
\label{sub-GeV}
\\
R_{\mu/e}^{\mathrm{multi-GeV}}
\null & \null = \null
0.57^{+0.08}_{-0.07} \pm 0.07
\label{multi-GeV}
\end{align}
indicate
an anomalous flavor composition in the 
observed atmospheric neutrino flux,
in both the sub-GeV and multi-GeV energy ranges.

In our analysis we did not use the ratio of ratios
$R_{\mu/e}$,
because,
according  to  the 
arguments discussed in Ref.\cite{Fogli2},
its non-Gaussian probability
distribution might bias the statistical analysis.
Therefore,
we preferred to study the separate $e$-like and $\mu$-like event rates, 
including in our analysis the correlation of their uncertainties.

\subsection{Theoretical calculation}

The expected numbers of $e$-like and $\mu$-like
events in each angular bin, under
the hypothesis of neutrino oscillations,
are given by
\begin{eqnarray}
&&
N_{e i}
=
N_{e i}^{\mathrm{th}}
\,
P^{(i)}_{\nu_{e}\to\nu_{e}}
+ 
N_{\mu i}^{\mathrm{th}}
\,
P^{(i)}_{\nu_{\mu}\to\nu_{e}}
\; ,
\label{3011}
\\
&&
N_{\mu i}
=
N_{e i}^{\mathrm{th}}
\,
P^{(i)}_{\nu_{e}\to\nu_{\mu}}
+
N_{\mu i}^{\mathrm{th}}
\,
P^{(i)}_{\nu_{\mu}\to\nu_{\mu}}
\,,
\label{3012}
\end{eqnarray}
where $ N_{e i}^{\mathrm{th}} $ and $ N_{\mu i}^{\mathrm{th}} $ 
are, respectively, the number of $e$-like and $\mu$-like events in 
the angular bin $i$ calculated theoretically,
under the assumption that 
there are no oscillations, while
$P^{(i)}_{\nu_{\alpha}\to\nu_{\beta}}$
is the averaged probability of
$\nu_{\alpha}\to\nu_{\beta}$ 
transitions
(with $ \alpha , \beta = e , \mu $)
in the bin $i$. 
The five bins of $\cos\theta$
for the Kamiokande multi-GeV data
are centered around
$ \left\langle \cos\theta \right\rangle_{i=1,\ldots,5} =
-0.8 \, , \, -0.4 \, , \, 0.0 \, ,$ $0.4 \, , \, 0.8 $,
which correspond to the average distances
$ \left\langle L \right\rangle_{i=1,\ldots,5} =
10230 \, , \, 5157 \, , \, 852 \, , \, 54 \, , \, 26 \, \mathrm{Km} $
travelled by neutrinos
from the production point down to the detector site.

Since
the neutrino oscillation probabilities
$P_{\nu_{\alpha}\to\nu_{\beta}}$
are functions of the neutrino energy and
of the zenith angle,
in order to
compare the theoretical calculation
with the experimental data, they must be 
averaged both on the neutrino energy spectrum
(different for $e$-like and $\mu$-like events)
and on the angular width of each 
zenith-angle bin used in our analysis.
 
Therefore,
a preliminary step  in  our calculations
consisted  of  the evaluation 
of the averaged probabilities
$
P^{(i)}_{\nu_{\alpha}\to\nu_{\beta}}
(m_{3}^2,|U_{e3}|^{2},|U_{\mu3}|^{2})
$
in each angular bin $i$,
for a fixed set of values of the three oscillation
parameters
$m_{3}^2$,
$|U_{e3}|^{2}$,
$|U_{\mu3}|^{2}$,
defined by the double-integrals
\begin{equation}
P^{(i)}_{\nu_{\alpha}\to\nu_{\beta}}(m_{3}^2,|U_{e3}|^{2},|U_{\mu3}|^{2})
=
\int \mathrm{d}\!\cos\theta
\int \mathrm{d}E \, S_{\nu_{\alpha}}(E) \, 
P_{\nu_{\alpha}\to\nu_{\beta}}
(E,\cos\theta;m_{3}^2,|U_{e3}|^{2},|U_{\mu3}|^{2})
\,,  
\label{302}
\end{equation}
where
$\theta$ is the zenith angle,
$E$ is the neutrino energy
and
$S_{\nu_{\alpha}}(E)$
is the energy spectrum
of the parent neutrino (with flavor $\alpha$).
These energy spectra were extracted from the figures published
in the Ref.\cite{KAM92}
for the Kamiokande sub-GeV data
and in Ref.\cite{KAM94}
for the Kamiokande multi-GeV data. 

When matter effects are included in the calculations,
the probabilities
$
P_{\nu_{\alpha}\to\nu_{\beta}}
(E,\cos\theta;m_{3}^2,|U_{e3}|^{2},|U_{\mu3}|^{2})
$
are not simple functions of the neutrino energy 
and zenith-angle,
but must be calculated numerically by
taking into account the matter density
along the neutrino path in
the Earth.
The density and composition of the interior of the Earth
is known from seismological measurements
(see Refs.\cite{Stacey,Anderson}).
The electron number density $N_{e}$
varies in a discontinuous way from
$ 1.6 \, N_A \, \mathrm{cm}^{-3} $
near the surface to
$ 6.2 \, N_A \, \mathrm{cm}^{-3} $
in the inner core.
However,
it is possible to approximate the electron number density
profile with a step-like function,
each step representing a shell with constant density
(see also Refs.\cite{Fogli1}).
We approximated the numerical data in Ref.\cite{Anderson}
with five shells of constant density
(see Fig.\ref{fig1})
and we considered
the atmosphere as an outer shell
with a thickness of
$ 20 \, \mathrm{Km} $.
The effective neutrino path in each of the shells
depends upon the zenith angle
of the neutrino trajectory.

The amplitude of
$ \nu_\alpha \to \nu_\beta $
transitions
for a neutrino with energy
$ E \simeq p $
in a medium with constant density
is given by 
Eq.(\ref{20219}).
In the case of a neutrino
crossing various shells of different density, 
the matrices describing the propagation
across different shells
must be multiplied serially,
as in Eq.(\ref{202161}),
in order to get the amplitude of 
the transition along the total path. 
The square modulus of this amplitude gives the probability of
$ \nu_\alpha \to \nu_\beta $
oscillations
(see Eq.(\ref{012})).
This probability must be folded
with the relevant neutrino energy spectrum
and integrated over 
the selected intervals of energy and $\cos\theta$
(see Eq.(\ref{302})).

We took the values of
$ N_{ei}^{\mathrm{th}} $ and $ N_{{\mu}i}^{\mathrm{th}} $
given by Table 2 (flux A) of Ref.\cite{KAM94}
for the Kamiokande sub-GeV data
and
by Fig.3 of Ref.\cite{KAM94} for the Kamiokande multi-GeV data. 
The estimated uncertainties of the absolute value of
$ N_{e i}^{\mathrm{th}} $ and $ N_{\mu i}^{\mathrm{th}} $
amount to 30\%,
mainly due to the 
uncertainties of the value of the
calculated electron and muon neutrino fluxes.
The estimated relative errors
for the ratios
$ N_{{\mu}i}^{\mathrm{th}} / N_{ei}^{\mathrm{th}} $
ratio are much smaller:
9\% for the Kamiokande sub-GeV data
and
12\% for the Kamiokande multi-GeV data.
In our calculations
we neglected the fact that not all
$e$-like and $\mu$-like events
are produced by
$ \nu_{e} $ ($\bar\nu_{e}$)
and
$\nu_{\mu} $ ($\bar\nu_{\mu}$)
interactions,
respectively.
The purity of the
$e$-like and $\mu$-like events
is estimated by the Kamiokande Collaboration
to be higher than 90\%
\cite{KAM94}.

\subsection{Results of the fit}
\label{Results of the fit}

The number of $e$-like and $\mu$-like events
was calculated using the formulas
(\ref{3011}) and  (\ref{3012})
for a grid of values of the parameters
$m_{3}^2$,
$|U_{e3}|^{2}$,
$|U_{\mu3}|^{2}$,
averaging over the contributions of neutrinos and anti-neutrinos.
These values were used together with the experimental 
results to build a proper $ \chi^2 $ function.
The best fit of the Kamiokande data
was obtained for
\begin{equation}
m_3^2 = 2.5 \times 10^{-2} \, \mathrm{eV}^2
\;,
\qquad
|U_{e3}|^2 = 0.26
\;,
\qquad
|U_{\mu3}|^2 = 0.49
\;,
\label{best-fit}
\end{equation}
with
$ \chi^2 = 6.9 $
for 9 degrees of freedom,
corresponding to a CL of 65\%. 

The regions allowed at 90\% CL by the Kamiokande data  
in the
$B_{{e};{e}}$--$m_{3}^2$,
$B_{{\mu};{\mu}}$--$m_{3}^2$,
$A_{{\mu};{e}}$--$m_{3}^2$ and
$A_{{\mu};{\tau}}$--$m_{3}^2$
planes
are presented
in
Figs.\ref{fig4}, \ref{fig5}, \ref{fig6} and \ref{fig7},
respectively
(shadowed regions).
Since the oscillation probabilities in vacuum
(\ref{2107})
have the same form
as the corresponding ones
in the case of two-neutrino mixing
(see Eqs.(\ref{21077})),
the allowed regions 
in the
$B_{{e};{e}}$--$m_{3}^2$,
$B_{{\mu};{\mu}}$--$m_{3}^2$,
$A_{{\mu};{e}}$--$m_{3}^2$ and
$A_{{\mu};{\tau}}$--$m_{3}^2$
planes
can be compared directly
with the exclusion plots
of the terrestrial reactor and accelerator
neutrino oscillation experiments
whose data have been analyzed by the experimental collaborations
under the assumption
of two-generation mixing.
This is simply done through the identification of the appropriate
$A_{{\alpha};{\beta}}$
or
$B_{{\alpha};{\alpha}}$
with
$\sin^22\vartheta$ 
and the identification of
$\Delta{m}^2$
with
$m_{3}^2$.
Moreover,
the expected sensitivities of future long-baseline experiments
are also presented
by the experimental collaborations
as curves in the
$\sin^22\vartheta$--$\Delta{m}^2$
plane relative to two-neutrino oscillations
in vacuum.
Therefore,
each of these sensitivity curves can be compared directly
with the corresponding allowed region in one of the
$A_{{\alpha};{\beta}}$--$m_{3}^2$
or
$B_{{\alpha};{\alpha}}$--$m_{3}^2$
planes.
The same reasoning applies to the future results
of reactor long-baseline experiments
\cite{CHOOZ,PaloVerde},
whose neutrino beams propagate in vacuum.
On the other hand,
the neutrino beams of accelerator long-baseline experiments
\cite{K2K,MINOS,ICARUS}
will propagate in the Earth
and matter effects may be significant.
Since the effects of matter are different
in the cases of two-neutrino and three-neutrino mixing,
in order to get information on the mixing parameters
in the scheme with
three neutrinos with a mass hierarchy
considered here
it will be necessary to analize the data
with the formalism presented in
Section \ref{Oscillations of 3 neutrinos}\footnote{
Let us notice that,
since the energy of accelerator neutrinos
in long-baseline experiments is of the order of 10 GeV,
the formalism presented in
Section \ref{Oscillations of 3 neutrinos}
can be applied without caveats also in the case
of the large mixing angle solution
\cite{SOLMSW,recent-sun}
of the solar neutrino problem,
for which
$ m_2^2 \lesssim 3 \times 10^{-5} \, \mathrm{eV}^2 $
and
$ m_2^2 R_{\oplus} / 2 p \lesssim 5 \times 10^{-2} $.
}.

The dashed and dotted curves in
Fig.\ref{fig4}
represent, respectively,
the exclusion curves obtained in the
Bugey \cite{Bugey}
and
Krasnoyarsk \cite{Krasnoyarsk}
reactor
$\bar\nu_e$
disappearance experiment
(the excluded region lie on the right of the curves).
One can see that a part of the shadowed region
allowed by the Kamiokande data
in the
$B_{{e};{e}}$--$m_{3}^2$
plane
is excluded by the results of the reactor
neutrino oscillation experiments
(the lightly shadowed region).
Only the darkly shadowed region
in Fig.\ref{fig4}
is allowed by the results of
the Kamiokande and reactor experiments.
In Fig.\ref{fig4}
we have also plotted the sensitivity curves
of the CHOOZ
\cite{CHOOZ}
and
Palo Verde
\cite{PaloVerde}
long-baseline
$\bar\nu_e$
disappearance experiments with reactor anti-neutrinos
(the dash-dotted and dash-dot-dotted curves).
One can see that a large part of the allowed region
will be explored by the
CHOOZ
and
Palo Verde
experiments.

The region allowed by the Kamiokande data
in the
$B_{{\mu};{\mu}}$--$m_{3}^2$
plane,
shown in Fig.\ref{fig5},
is not constrained by the results of terrestrial
neutrino oscillation experiments.
Indeed,
the best limit at small values of $\Delta{m}^2$
is provided by the
CDHS
\cite{CDHS84}
$
\stackrel{\makebox[0pt][l]
{$\hskip-3pt\scriptscriptstyle(-)$}}{\nu_{\mu}}
$
disappearance experiment,
whose exclusion curve
lies rather far from the Kamiokande-allowed region,
as shown in Fig.\ref{fig5}
(the dashed curve).
In Fig.\ref{fig4}
we have also plotted the sensitivity curves
of the
K2K \cite{K2K} and
MINOS \cite{MINOS}
long-baseline experiments
(the dash-dotted and dash-dot-dotted curves),
which show that the Kamiokande-allowed
region will be explored in the near future.

A part of the shadowed region
allowed by the Kamiokande data
in the
$A_{{\mu};{e}}$--$m_{3}^2$
plane,
shown in Fig.\ref{fig6},
is excluded by the results of the reactor
neutrino oscillation experiments
(the lightly shadowed region),
leaving the darkly shadowed region allowed
by the results
of the Kamiokande and reactor experiments.
As shown in Fig.\ref{fig6},
most of this allowed region will be explored by the
K2K,
MINOS and
ICARUS \cite{ICARUS}
long-baseline experiments,
whose sensitivity curves
are represented by the
long-dashed, dash-dotted and dash-dot-dotted curves,
respectively.

Finally,
Fig.\ref{fig7} shows the Kamiokande-allowed region
in the
$A_{{\mu};{\tau}}$--$m_{3}^2$
plane,
part of which
is indirectly excluded by the results of the reactor
neutrino oscillation experiments
(the lightly shadowed region).
Indeed,
in the lightly shadowed region
the amplitude
$ A_{\nu_{\mu};\nu_{\tau}} $
is rather small and
the fit
of the Kamiokande data requires large values
of the amplitude
$ A_{\nu_{\mu};\nu_{e}} $,
which,
because of the inequality
$ A_{\nu_{\mu};\nu_{e}} \le B_{\nu_{e};\nu_{e}} $,
lie in the region excluded by
the results of the Bugey and Krasnoyarsk
reactor experiments.
Hence,
only the darkly shadowed region
is allowed by the results of the
Kamiokande and reactor experiments.
As shown in Fig.\ref{fig7},
this region will be explored by
the MINOS and ICARUS
long-baseline experiments,
whose sensitivity curves
are represented by the
dash-dotted and dash-dot-dotted curves,
respectively.
Since the allowed region extends
at large values of
$ A_{\nu_{\mu};\nu_{\tau}} $,
the Kamiokande data indicate that
$\nu_\mu\to\nu_\tau$
oscillations could be observed
with a large statistics by the
MINOS and ICARUS
experiments.

The results of our analysis
are similar to those presented
in Ref.\cite{BGK},
where the same scheme with three neutrinos
and the mass hierarchy (\ref{pattern}) was used, 
but matter effects were not taken into account.
However,
one must notice that
the presence of matter
is important because it modifies the phases
of neutrino oscillations
\cite{Pantaleone94}
and its effect is to slightly enlarge
the allowed region of the mixing parameters
towards low values of
$ m_3^2 $.

\section{Conclusions}
\label{Conclusions}

We have presented a comprehensive
formalism
for the description of neutrino oscillations
in the Earth
in the case with three massive neutrinos
whose masses satisfy the hierarchical pattern
(\ref{pattern}).
Such a scheme
is allowed by the see-saw mechanism
for the generation of neutrino masses
and
permits one to explain
the solar and atmospheric neutrino problems
through neutrino oscillations if
$
m_2^2
\sim
10^{-5} \, \mbox{or} \, 10^{-10} \, \mathrm{eV}^2
$
and
$
m_3^2
\sim
10^{-2} \, \mathrm{eV}^2
$.

In Section \ref{Low energy neutrinos}
we have discussed the matter effect
on the oscillations of neutrinos with low energy
and  have shown that
the oscillation probabilities depend on two oscillation lengths:
$L_{\mathrm{osc}}^{\mathrm{short}}$
which is the same as the oscillation length in vacuum
and
$
L_{\mathrm{osc}}^{\mathrm{long}}
\sim
10^{4} \, \mathrm{Km}
$
(see Eq.(\ref{203061})).
The oscillation length
$L_{\mathrm{osc}}^{\mathrm{long}}$
is independent of  the value of
the mass $m_3$ of the heavy neutrino
and  the neutrino energy,
and is very sensitive to the matter density.
Future very-long-baseline
neutrino oscillation experiments
could observe the oscillations due to
$L_{\mathrm{osc}}^{\mathrm{long}}$, after
averaging  the oscillation probabilities
over the fast oscillations due to
$L_{\mathrm{osc}}^{\mathrm{short}}$.
The sensitivity of these averaged
oscillations to the matter density
along the neutrino path
could allow one to obtain
a detailed tomography of the interior
of the Earth.

In Section \ref{Analysis}
we discussed the analysis of
the Kamiokande atmospheric data
and
 presented our results
in the form of allowed regions in the
$B_{{e};{e}}$--$m_{3}^2$,
$B_{{\mu};{\mu}}$--$m_{3}^2$,
$A_{{\mu};{e}}$--$m_{3}^2$ and
$A_{{\mu};{\tau}}$--$m_{3}^2$
planes
(see
Figs.\ref{fig4}, \ref{fig5}, \ref{fig6} and \ref{fig7}).
The knowledge of the allowed regions
in these planes is useful because
the oscillation probabilities in vacuum
(\ref{2107})
in the scheme under consideration
have the same form
as the oscillation probabilities in vacuum
(\ref{21077})
in the case of two-neutrino mixing,
which have been used by the experimental collaborations in the analysis
of the data of terrestrial
oscillation experiments
with reactor and accelerator neutrinos,
yielding exclusion curves
in the
$\sin^{2}2\vartheta$--$\Delta{m}^{2}$
plane.
Identifying the appropriate
$A_{\alpha;\beta}$
or
$B_{\alpha;\alpha}$
with
$\sin^{2}2\vartheta$
and
$m_3^2$
with
$\Delta{m}^{2}$,
the exclusion curves in the
$\sin^{2}2\vartheta$--$\Delta{m}^{2}$
plane
obtained by the terrestrial
neutrino oscillation experiments
can be plotted directly
in the
$B_{{e};{e}}$--$m_{3}^2$,
$B_{{\mu};{\mu}}$--$m_{3}^2$,
$A_{{\mu};{e}}$--$m_{3}^2$ and
$A_{{\mu};{\tau}}$--$m_{3}^2$
planes
and can be compared directly with the
regions allowed by the analysis of the Kamiokande data.
The same reasoning applies to
the sensitivity curves
and future results of long-baseline experiments
evaluated under the assumption of two-neutrino oscillations in vacuum.

Our analysis of the Kamiokande data
took into account the presence of matter
in the oscillations of neutrinos
passing through the Earth,
whose effect is to slightly enlarge the allowed
region towards low values of
$m_3^2$.
The best fit is obtained for
$ m_3^2 = 2.5 \times 10^{-2} \, \mathrm{eV}^2 $
and from 
Figs.\ref{fig4}, \ref{fig5}, \ref{fig6} and \ref{fig7}
one can see that the allowed region of
$m_3^2$
extends from about
$ 3.5 \times 10^{-3} \, \mathrm{eV}^2 $
to about
$ 3.5 \times 10^{-2} \, \mathrm{eV}^2 $.

From
Figs.\ref{fig4}, \ref{fig5}, \ref{fig6} and \ref{fig7}
one can see that the
long-baseline experiments
with reactor
(CHOOZ and Palo Verde)
and accelerator
(K2K, MINOS and ICARUS)
neutrinos
could observe neutrino oscillations
in all channels
with a relatively large statistics\footnote{After we finished this work
the CHOOZ collaboration disclosed its first results
(M. Apollonio \textit{et al.},
Phys. Lett. B \textbf{420}, 397 (1998)),
which are compatible
with the hypothesis of absence of
$\bar\nu_e$
oscillations
and lead to the upper bound
$ \sin^22\vartheta \lesssim 0.18 $
for
$ \Delta{m}^2 \gtrsim 10^{-3} \, \mathrm{eV}^2 $
at 90\% CL.
This limit is close to the CHOOZ sensitivity curve
shown in Fig.\ref{fig4}
and implies that large
$\nu_\mu\leftrightarrows\nu_e$
oscillations of atmospheric neutrinos
are excluded.
However,
a subdominant contribution from
$\nu_\mu\leftrightarrows\nu_e$
oscillations
to the atmospheric neutrino anomaly
remains an interesting possibility
(see also
G.L. Fogli, E. Lisi, A. Marrone and D. Montanino,
hep-ph/9711421).}.

The region in the
$A_{{\mu};{\tau}}$--$m_{3}^2$
plane
allowed by the
results of the Kamiokande
and reactor neutrino oscillation experiments
extends
at large values of
$ A_{\nu_{\mu};\nu_{\tau}} $
(see Fig.\ref{fig7}).
Hence,
the Kamiokande data indicate that
the
MINOS and ICARUS
long-baseline experiments
could observe a relatively large signal
in the channel
$\nu_\mu\to\nu_\tau$.
This could be the first direct observation
of the tau neutrino.

\begin{flushleft}
\Large \textbf{Acknowledgments}
\end{flushleft}

We would like to thank
J. Pantaleone for his suggestion of
the effect discussed
in Section \ref{Low energy neutrinos}.

\appendix

\section{The mixing matrix}
\label{The mixing matrix}

A $ 3 \times 3 $
unitary matrix
$U$
can be written as
(see Refs.\cite{Murnaghan62,SV80})
\begin{equation}
U
=
D(\omega)
\,
\prod_{a<b}
W_{ab}\!\left(\vartheta_{ab}e^{i\eta_{ab}}\right)
\qquad
(a,b=1,2,3)
\,,
\label{100001}
\end{equation}
with the unitary matrices
\begin{align}
\null & \null
D(\omega)
=
\mathrm{diag}\!\left(
e^{i\omega_{1}}
\, , \,
e^{i\omega_{2}}
\, , \,
e^{i\omega_{3}}
\right)
\,,
\label{100002}
\\
\null & \null
\left[
W_{ab}\!\left(\vartheta_{ab}e^{i\eta_{ab}}\right)
\right]_{rs}
=
\delta_{rs}
+
\left( \cos\vartheta_{ab} - 1 \right)
\left(
\delta_{ra} \, \delta_{sa}
+
\delta_{rb} \, \delta_{sb}
\right)
\nonumber
\\
\null & \null
\hskip+5cm
+
\sin\vartheta_{ab}
\left(
e^{i\eta_{ab}} \, \delta_{ra} \, \delta_{sb}
-
e^{-i\eta_{ab}} \, \delta_{rb} \, \delta_{sa}
\right)
\,.
\label{100003}
\end{align}
Here
$D(\omega)$
is a diagonal matrix
depending from the set of phases
$\omega=(\omega_{1},\omega_{2},\omega_{3})$
and
the matrices
$W_{ab}\!\left(\vartheta_{ab}e^{i\eta_{ab}}\right)$
are
unitary and unimodular.
For example,
we have
\begin{equation}
W_{12}(\vartheta_{12}e^{i\eta_{12}})
=
\begin{pmatrix}
\cos\vartheta_{12} & \sin\vartheta_{12} \, e^{i\eta_{12}} & 0
\\
-\sin\vartheta_{12} \, e^{-i\eta_{12}} & \cos\vartheta_{12} & 0
\\
0 & 0 & 1
\end{pmatrix}
\,.
\label{100004}
\end{equation}
The matrices
$D(\lambda)$
and
$W_{ab}\!\left(\vartheta_{ab}e^{i\eta_{ab}}\right)$
satisfy the useful identity
\begin{equation}
D(\lambda)
\,
W_{ab}\!\left(\vartheta_{ab}e^{i\eta_{ab}}\right)
D(\lambda)^{\dagger}
=
W_{ab}\!\left(
\vartheta_{ab}e^{i(\lambda_{a}+\eta_{ab}-\lambda_{b})}
\right)
\,,
\label{100006}
\end{equation}
for any set of phases
$\lambda=(\lambda_{1},\lambda_{2},\lambda_{3})$.
Using the identity (\ref{100006})
it is clear that the matrix
$W_{ab}\!\left(\vartheta_{ab}e^{i\eta_{ab}}\right)$
can be written as
\begin{equation}
W_{ab}\!\left(\vartheta_{ab}e^{i\eta_{ab}}\right)
=
D_{ab}
\,
V_{ab}
\,
D_{ab}^{\dagger}
\,,
\label{1000031}
\end{equation}
with
\begin{align}
\null & \null
\left[
D_{ab}
\right]_{rs}
=
\delta_{rs}
+
\left( e^{i\eta_{ab}} - 1 \right)
\delta_{ra} \, \delta_{sa}
\,,
\label{1000032}
\\
\null & \null
\left[
V_{ab}
\right]_{rs}
=
\delta_{rs}
+
\left( \cos\vartheta_{ab} - 1 \right)
\left( \delta_{ra} \, \delta_{sa} + \delta_{rb} \, \delta_{sb} \right)
+
\sin\vartheta_{ab}
\left(
\delta_{ra} \, \delta_{sb}
-
\delta_{rb} \, \delta_{sa}
\right)
\,.
\label{1000033}
\end{align}
The matrix
$V_{ab}$
operates
a rotation of an angle
$\vartheta_{ab}$
in the
$ \nu_{a} $--$ \nu_{b} $
plane.
For example,
we have
\begin{equation}
V_{12}
=
\begin{pmatrix}
\cos\vartheta_{12} & \sin\vartheta_{12} & 0
\\
-\sin\vartheta_{12} & \cos\vartheta_{12} & 0
\\
0 & 0 & 1
\end{pmatrix}
\,,
\qquad
D_{12}
=
\begin{pmatrix}
e^{i\eta_{12}} & 0 & 0
\\
0 & 1 & 0
\\
0 & 0 & 1
\end{pmatrix}
\,.
\label{1000041}
\end{equation}

The expression (\ref{100001}) for $U$
can be written as
\begin{equation}
U
=
D(\omega-\lambda)
\left(
\prod_{a<b}
D(\lambda)
W_{ab}\!\left(\vartheta_{ab}e^{i\eta_{ab}}\right)
D(\lambda)^{\dagger}
\right)
D(\lambda)
\,,
\label{100005}
\end{equation}
with the set of arbitrary phases
$\lambda=(\lambda_{1},\lambda_{2},\lambda_{3})$.
Using the identity (\ref{100006}),
we have
\begin{equation}
U
=
D(\omega-\lambda)
\left(
\prod_{a<b}
W_{ab}\!\left(
\vartheta_{ab}e^{i(\lambda_{a}+\eta_{ab}-\lambda_{b})}
\right)
\right)
D(\lambda)
\,.
\label{100007}
\end{equation}
The set of arbitrary phases
$\lambda=(\lambda_{1},\lambda_{2},\lambda_{3})$
can be chosen in order to
extract two of the three phases
$\eta_{ab}$
from the corresponding
$W_{ab}$.
Only two phases $\eta_{ab}$ can be extracted
because there are only two independent differences
$ \lambda_{a}-\lambda_{b} $.
For example,
one can extract
$\eta_{13}$
and
$\eta_{23}$
with the choice
\begin{equation}
\lambda_{1}-\lambda_{3}
=
- \eta_{13}
\,,
\qquad
\lambda_{2}-\lambda_{3}
=
- \eta_{23}
\,.
\label{1000071}
\end{equation}
Then,
$\eta_{12}$
cannot be extracted because
$
\lambda_{1}-\lambda_{2}
=
- \eta_{13} + \eta_{23}
$
is no longer arbitrary.

The order of the product of the
unitary matrices
$W_{ab}$
in Eq.(\ref{100001})
is arbitrary.
Making the choice (\ref{1000071})
and
$\lambda_{3}=0$,
the mixing matrix can be written as
\begin{equation}
U
=
D(\omega-\lambda)
\,
V_{23}
\,
V_{13}
\,
W_{12}
\,
D(\lambda)
\,.
\label{1000072}
\end{equation}
with
\begin{align}
\null & \null
D(\omega-\lambda)
=
\mathrm{diag}\!\left(
e^{i(\omega_{1}+\eta_{13})}
\, , \,
e^{i(\omega_{2}+\eta_{23})}
\, , \,
e^{i\omega_{3}}
\right)
\,,
\label{10000721}
\\
\null & \null
W_{12}
\equiv
W_{12}\!\left(
\vartheta_{12}e^{i\eta_{12}}
\right)
=
D_{12}
V_{12}
D_{12}^{\dagger}
\,,
\label{10000722}
\\
\null & \null
D(\lambda)
=
\mathrm{diag}\!\left(
e^{-i\eta_{13}}
\, , \,
e^{-i\eta_{23}}
\, , \,
1
\right)
\,.
\label{10000723}
\end{align}

The mixing matrix $U$
appears in the lepton charged-current
(the charged lepton fields
are defined as mass eigenstates,
because they are observed
in the experiments,
whereas neutrinos are never observed directly)
\begin{equation}
\begin{split}
j^{\mu}
\null & \null = \null
2
\sum_{\alpha=e,\mu,\tau}
\ell_{{\alpha}L}
\,
\gamma^{\mu}
\,
\nu_{{\alpha}L}
=
2
\sum_{\alpha=e,\mu,\tau}
\sum_{k=1,2,3}
\ell_{{\alpha}L}
\,
\gamma^{\mu}
\,
U_{{\alpha}k}
\,
\nu_{kL}
\\
\null & \null = \null
2
\sum_{\alpha,k}
\ell_{{\alpha}L}
\,
\gamma^{\mu}
\,
\left[
D(\omega-\lambda)
\,
V_{23}
\,
V_{13}
\,
W_{12}
\,
D(\lambda)
\right]_{{\alpha}k}
\,
\nu_{kL}
\,,
\end{split}
\label{100008}
\end{equation}
where
$\ell_{{\alpha}L}$
represents the left-handed component
of the charged lepton field
$\ell_{{\alpha}}$
(with $\alpha=e,\mu,\tau$).
The three phases in
$D(\omega-\lambda)$
can be eliminated
with a redefinition of the
arbitrary phases of the charged lepton fields,
$
\ell_{\alpha}
\to
\ell_{\alpha}
\,
e^{-i(\omega_{\alpha}-\lambda_{\alpha})}
$,
which leads to
\begin{equation}
j^{\mu}
=
2
\sum_{\alpha,k}
\ell_{{\alpha}L}
\,
\gamma^{\mu}
\,
\left[
V_{23}
\,
V_{13}
\,
W_{12}
\,
D(\lambda)
\right]_{{\alpha}k}
\,
\nu_{kL}
\,.
\label{100009}
\end{equation}
If neutrinos are Dirac particles,
also the phases of the neutrino fields
are arbitrary
and
the two phases in
$D(\lambda)$
can be eliminated
with the redefinition
$
\nu_{k}
\to
e^{-i\lambda_{k}}
\,
\nu_{k}
$.
In the case of
Majorana neutrinos
the elimination of the two phases
contained in $D(\lambda)$
(called in this case ``Majorana phases'')
is not possible,
because the Majorana mass term
is not invariant under rephasing
of the neutrino fields\footnote{
For instance,
the Majorana mass term
for one left-handed neutrino field $\nu_{L}$
is proportional to
$
\overline{\nu_{L}} \nu_{L}^{c}
+
\overline{\nu^{c}_{L}} \nu_{L}
=
- \nu_{L}^{\dagger} \mathcal{C} \nu_{L}^{*}
- \nu_{L}^{T} \mathcal{C}^{\dagger} \nu_{L}
$,
where
$ \nu_{L}^{c} = \mathcal{C} \overline{\nu}_{L}^{T} $
and
$\mathcal{C}$
is the charge conjugation matrix.
It is clear that the Majorana mass term
is not invariant
under the rephasing
$ \nu \to e^{-i\lambda} \nu $.
}.
However,
the presence of the Majorana phases
does not have any effect on neutrino oscillations
\cite{BHP80,Doi81,Langacker87}
(see Eq.(\ref{0163})).
In conclusion,
the general expression for the
mixing matrix
can be written as
\begin{equation}
U
=
V_{23}
\,
V_{13}
\,
W_{12}
\,
D(\lambda)
\,.
\label{100010}
\end{equation}
Explicitly we have
\begin{equation}
U
=
\begin{pmatrix}
c_{12}
c_{13}
e^{i\lambda_{1}}
&
s_{12}
c_{13}
e^{i(\eta_{12}+\lambda_{2})}
&
s_{13}
\\
-
s_{12}
c_{23}
e^{-i(\eta_{12}-\lambda_{1})}
-
c_{12}
s_{23}
s_{13}
e^{i\lambda_{1}}
&
c_{12}
c_{23}
e^{i\lambda_{2}}
-
s_{12}
s_{23}
s_{13}
e^{i(\eta_{12}+\lambda_{2})}
&
s_{23}
c_{13}
\\
s_{12}
s_{23}
e^{-i(\eta_{12}-\lambda_{1})}
-
c_{12}
c_{23}
s_{13}
e^{i\lambda_{1}}
&
-
c_{12}
s_{23}
e^{i\lambda_{2}}
-
s_{12}
c_{23}
s_{13}
e^{i(\eta_{12}+\lambda_{2})}
&
c_{23}
c_{13}
\end{pmatrix}
\,,
\label{100011}
\end{equation}
with
$ c_{ij} \equiv \cos\vartheta_{ij} $
and
$ s_{ij} \equiv \sin\vartheta_{ij} $.
This parameterization of the mixing matrix
is useful for the study
of neutrino oscillations in matter
(see Section \ref{Oscillations of 3 neutrinos}).

\newpage

\begin{flushleft}
\Large \textbf{Figure Captions}
\end{flushleft}
\begin{description}

\item[Figure~\ref{fig1}.]
\refstepcounter{figure} \label{fig1}
Density $\rho$
(A)
and
electron number density
(B)
in the interior of the Earth
as functions of the radial distance $r$
from the center of the Earth.
The solid curves represent
the data given in Ref.\cite{Anderson}.
The electron number density
is given by
$ N_e = N_A (\rho/\mathrm{g}) \langle Z / A \rangle $
with
$ \langle Z / A \rangle = 0.475 $
for
$ r \leq 3480 \, \mathrm{Km} $
(core)
and
$ \langle Z / A \rangle = 0.495 $
for
$ r > 3480 \, \mathrm{Km} $
(mantle).
The dotted curves represent
our approximation in terms of five shells
with outer radii
$
r_{i=1,\ldots,5}
=
1221,3480,5701,5971,6371
\, \mathrm{Km}
$,
having constant densities
$
\rho_{i=1,\ldots,5}
=
13.0,11.3,5.0,3.9,3.0
\, \mathrm{g}/\mathrm{cm}^3
$
and
electron number density
$
(N_e)_{i=1,\ldots,5}
=
6.15,5.36,2.47,1.93,1.50
\, N_A \, \mathrm{cm}^{-3}
$.

\item[Figure~\ref{fig2}.]
\refstepcounter{figure} \label{fig2}
Averaged value
(\ref{2021633})
of the survival probabilities of muon neutrinos
(solid curves)
and anti-neutrinos (dashed curves)
as functions of the distance $L=t$
in a medium with constant electron density
$ N_{e} = 3 \, N_A \, \mathrm{cm}^{-3} $.
Two sets of values of the mixing parameters
are considered:
(A)
$|U_{e3}|^2=|U_{\mu3}|^2=1/3$
(which corresponds to maximal mixing
of the three neutrinos)
and
(B)
$|U_{e3}|^2=0.3$,
$|U_{\mu3}|^2=0.5$
(which is close to the best fit
of the Kamiokande data;
see Eq.(\ref{best-fit})).
The dotted lines represent the
averaged survival probabilities in vacuum
(\ref{20216331}),
which are equal for neutrinos and anti-neutrinos.

\item[Figure~\ref{fig3}.]
\refstepcounter{figure} \label{fig3}
Fig.\ref{fig3}A,B:
$
\left\langle
P_{\nu_\mu\to\nu_\mu}
\right\rangle
$
(solid curves)
and
$
\left\langle
P_{\bar\nu_\mu\to\bar\nu_\mu}
\right\rangle
$
(dashed curves)
for neutrinos propagating
in the interior of the Earth
as a function of the zenithal angle $\theta$.
These figures
correspond to the
same two sets of values of the mixing parameters
as in Fig.\ref{fig2}.
The dotted lines represent the
averaged survival probabilities in vacuum
(\ref{20216331}),
which are equal for neutrinos and anti-neutrinos.
Fig.\ref{fig3}C:
slant depth
as a function of the zenithal angle $\theta$.

\item[Figure~\ref{fig4}.]
\refstepcounter{figure} \label{fig4}
Region in the
$B_{{e};{e}}$--$m_{3}^2$
plane
allowed at 90\% CL by the Kamiokande data
(the shadowed region).
The lightly shadowed region
is forbidden by the results of the
Bugey \cite{Bugey}
and
Krasnoyarsk \cite{Krasnoyarsk}
reactor
$\bar\nu_e$
disappearance experiments,
whose exclusion curves
are represented by the
dahed and dotted lines, respectively,
The dash-dotted and dash-dot-dotted curves
represent the sensitivity
of the CHOOZ \cite{CHOOZ}
and
Palo Verde \cite{PaloVerde}
reactor long-baseline
experiments.

\item[Figure~\ref{fig5}.]
\refstepcounter{figure} \label{fig5}
Region allowed at 90\% CL by the Kamiokande data
in the
$B_{{\mu};{\mu}}$--$m_{3}^2$
plane
(the shadowed region).
The dashed line reproduces the exclusion curve of the
CDHS \cite{CDHS84}
$
\stackrel{\makebox[0pt][l]
{$\hskip-3pt\scriptscriptstyle(-)$}}{\nu_{\mu}}
$
disappearance experiment.
The dash-dotted and dash-dot-dotted curves
represent the sensitivity
of the
K2K \cite{K2K} and
MINOS \cite{MINOS}
accelerator long-baseline experiments.

\item[Figure~\ref{fig6}.]
\refstepcounter{figure} \label{fig6}
Region in the
$A_{{\mu};{e}}$--$m_{3}^2$
plane
allowed at 90\% CL by the
analysis of the Kamiokande data
(the shadowed region).
The lightly shadowed region is excluded
by the results
of the
Bugey \cite{Bugey}
and
Krasnoyarsk \cite{Krasnoyarsk}
reactor neutrino experiments.
The long-dashed, dash-dotted and dash-dot-dotted curves
represent the sensitivities of the
K2K \cite{K2K},
MINOS \cite{MINOS} and
ICARUS \cite{ICARUS}
long-baseline experiments.

\item[Figure~\ref{fig7}.]
\refstepcounter{figure} \label{fig7}
Region
in the
$A_{{\mu};{\tau}}$--$m_{3}^2$
plane
allowed at 90\% CL by the analysis of the Kamiokande data
(the shadowed region).
The lightly shadowed region
is indirectly excluded by the results of the reactor
neutrino oscillation experiments,
leaving the darkly shadowed allowed region.
The dash-dotted and dash-dot-dotted curves
represent the sensitivities of
the MINOS and ICARUS
long-baseline experiments
in the $\nu_\mu\to\nu_\tau$ channel.

\end{description}


\begin{figure}[p]
\null\vspace{-1cm}
\begin{center}
\includegraphics[bb=80 110 540 780,width=0.9\textwidth]{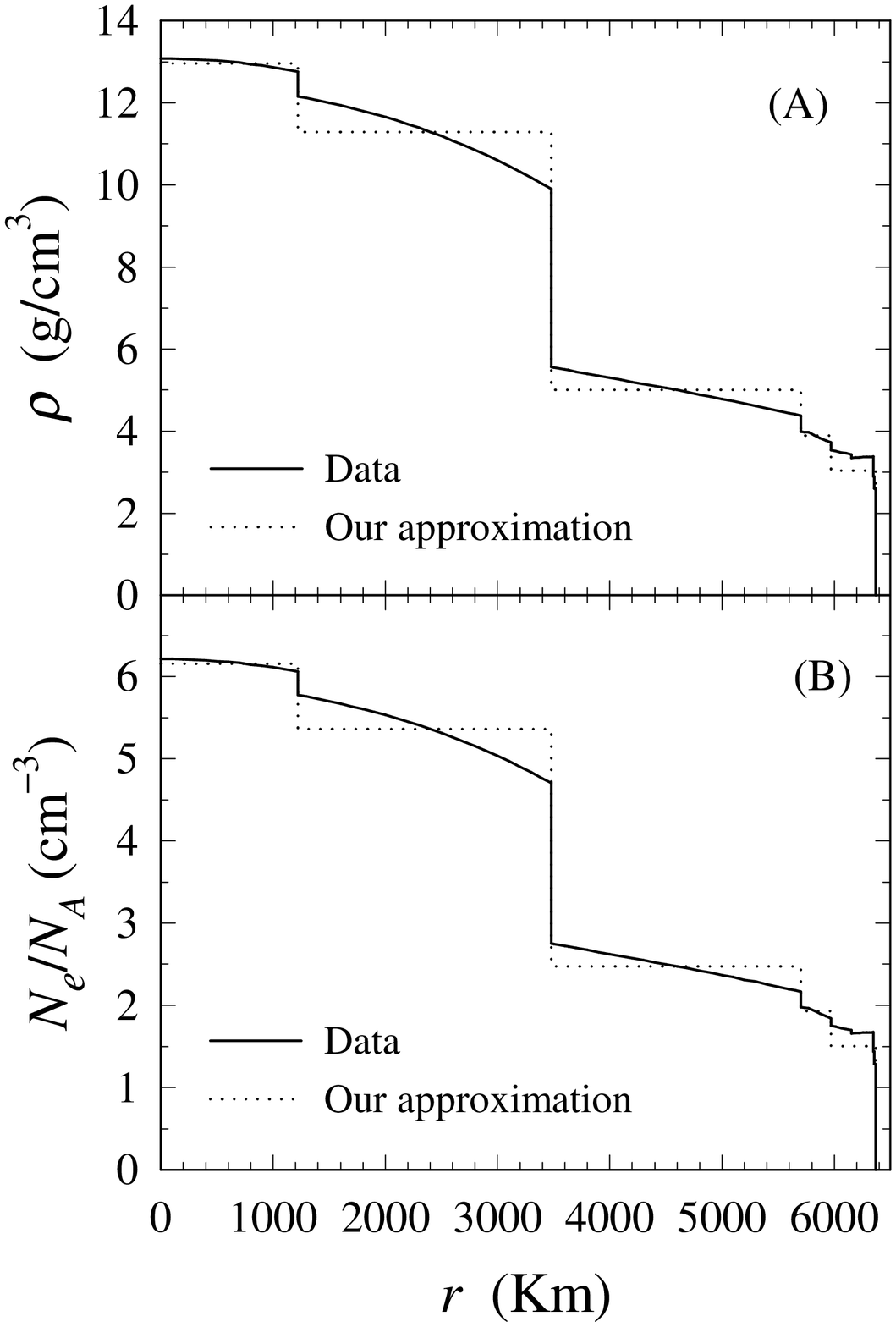}
\\[0.5cm]
Figure~\ref{fig1}
\end{center}
\end{figure}

\begin{figure}[p]
\null\vspace{-1cm}
\begin{center}
\includegraphics[bb=60 110 540 780,width=0.9\textwidth]{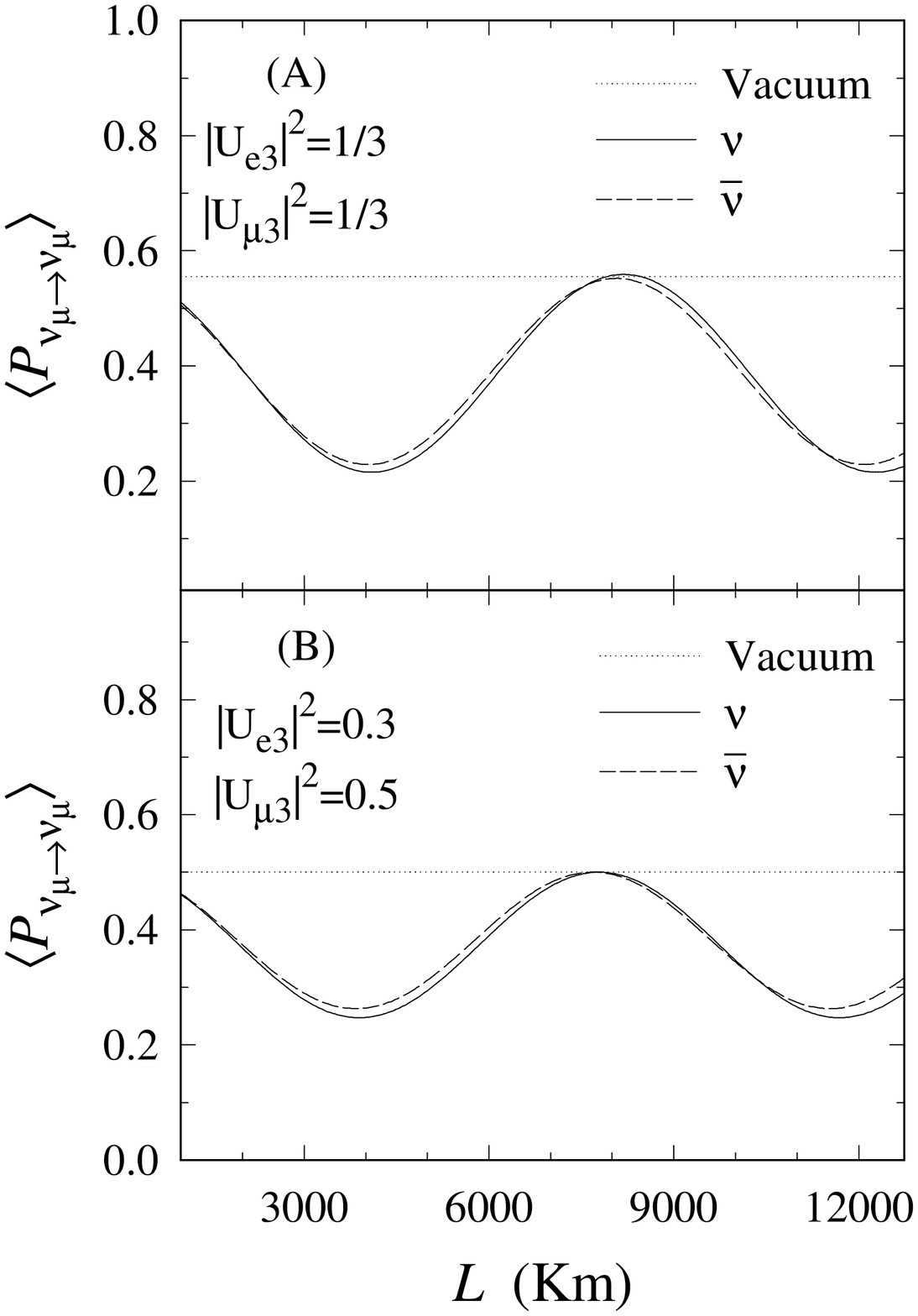}
\\[0.5cm]
Figure~\ref{fig2}
\end{center}
\end{figure}

\begin{figure}[p]
\null\vspace{-1cm}
\begin{center}
\includegraphics[bb=60 110 540 780,width=0.9\textwidth]{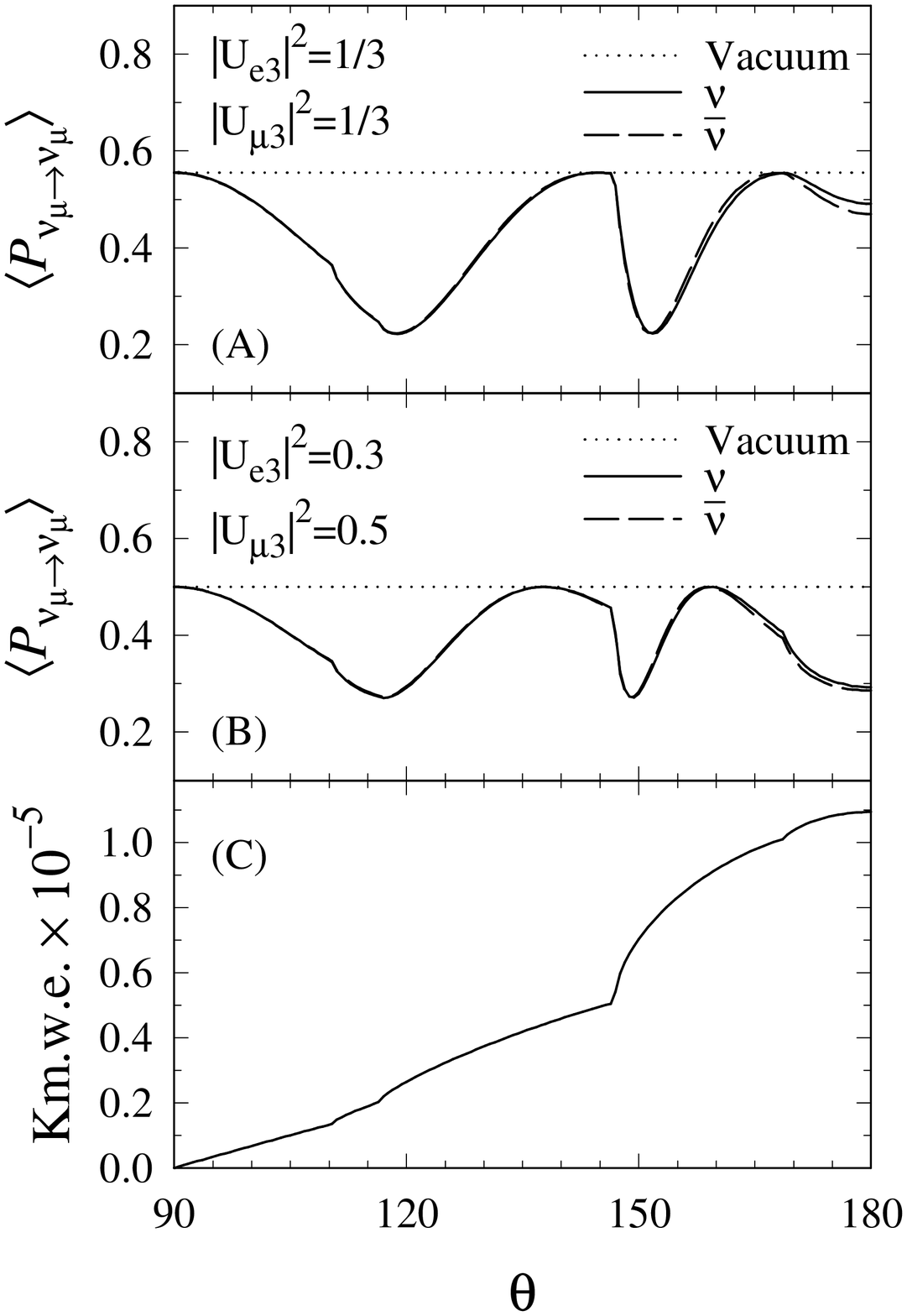}
\\[0.5cm]
Figure~\ref{fig3}
\end{center}
\end{figure}

\begin{figure}[p]
\null\vspace{-1cm}
\begin{center}
\includegraphics[bb=30 220 550 740,width=0.9\textwidth]{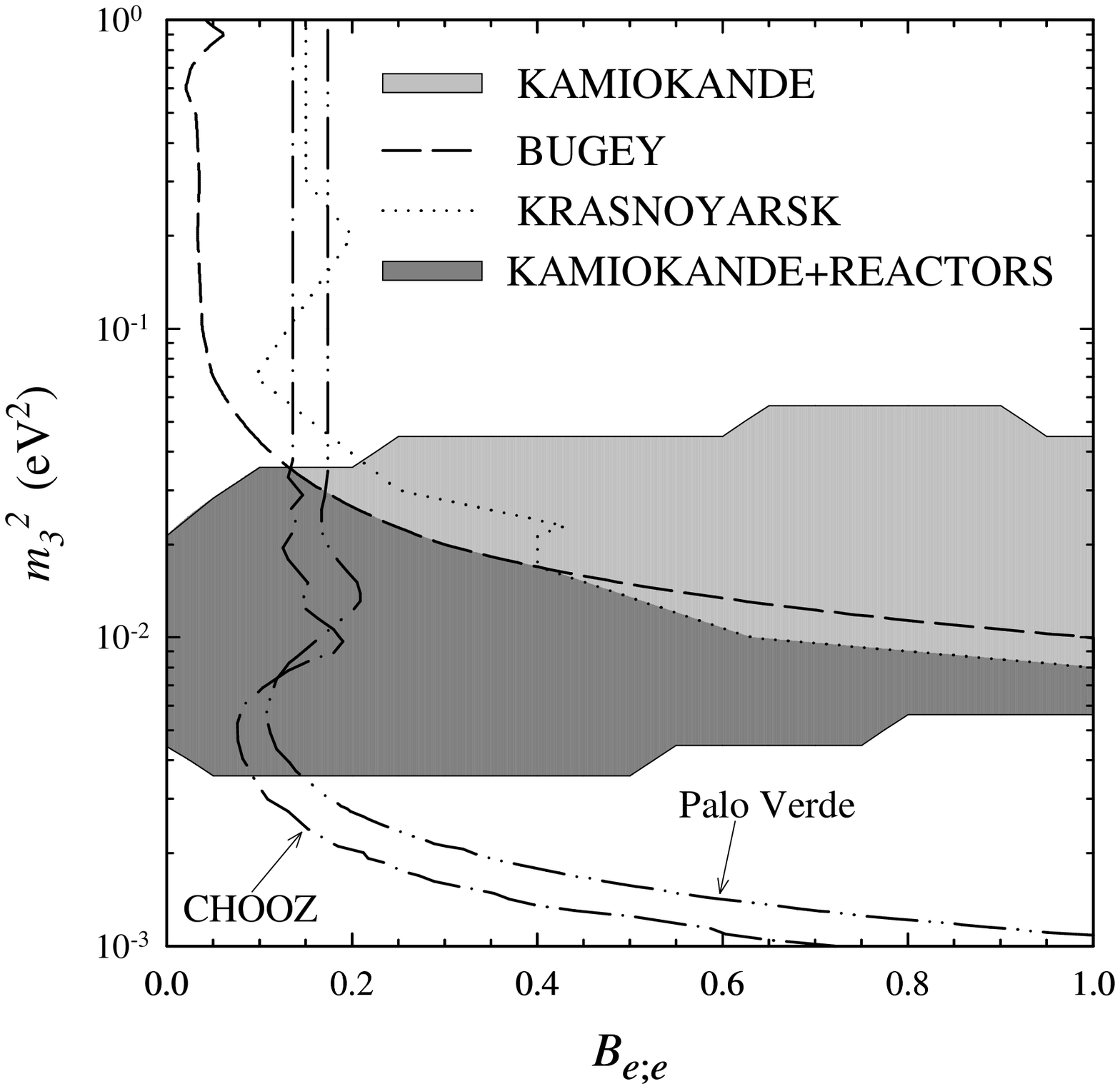}
\\[0.5cm]
Figure~\ref{fig4}
\end{center}
\end{figure}

\begin{figure}[p]
\null\vspace{-1cm}
\begin{center}
\includegraphics[bb=30 220 550 740,width=0.9\textwidth]{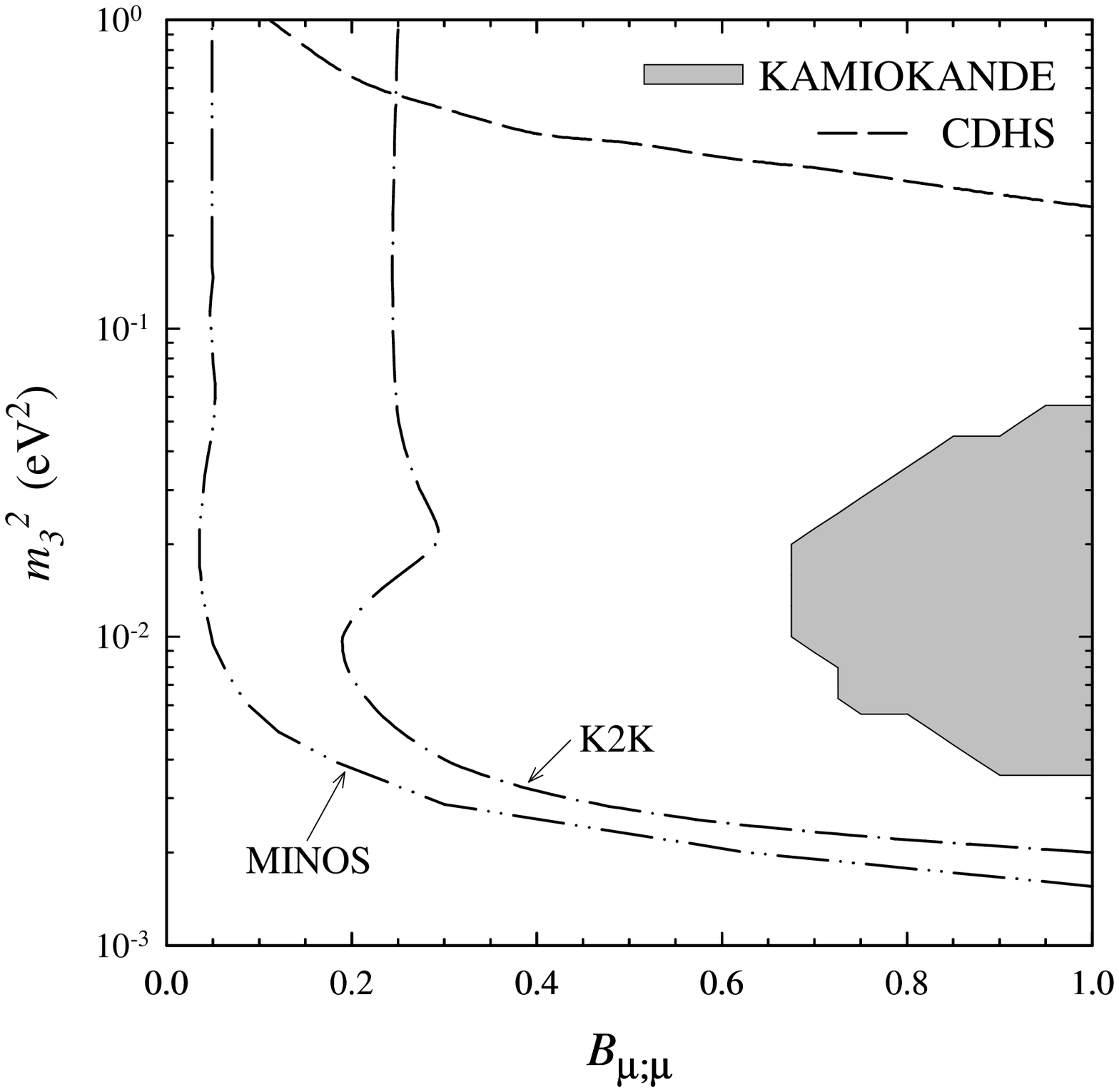}
\\[0.5cm]
Figure~\ref{fig5}
\end{center}
\end{figure}

\begin{figure}[p]
\null\vspace{-1cm}
\begin{center}
\includegraphics[bb=30 220 550 740,width=0.9\textwidth]{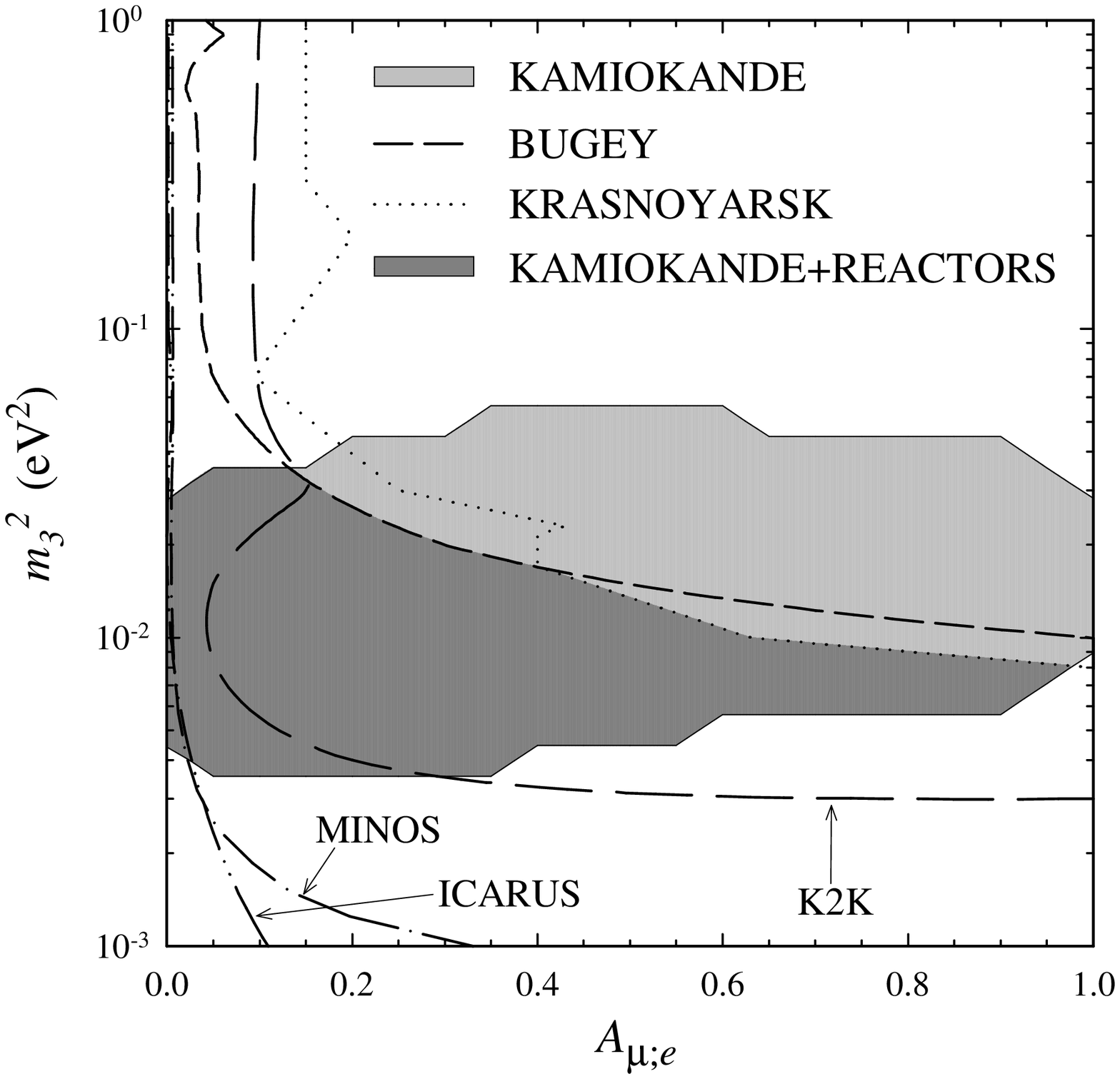}
\\[0.5cm]
Figure~\ref{fig6}
\end{center}
\end{figure}

\begin{figure}[p]
\null\vspace{-1cm}
\begin{center}
\includegraphics[bb=30 220 550 740,width=0.9\textwidth]{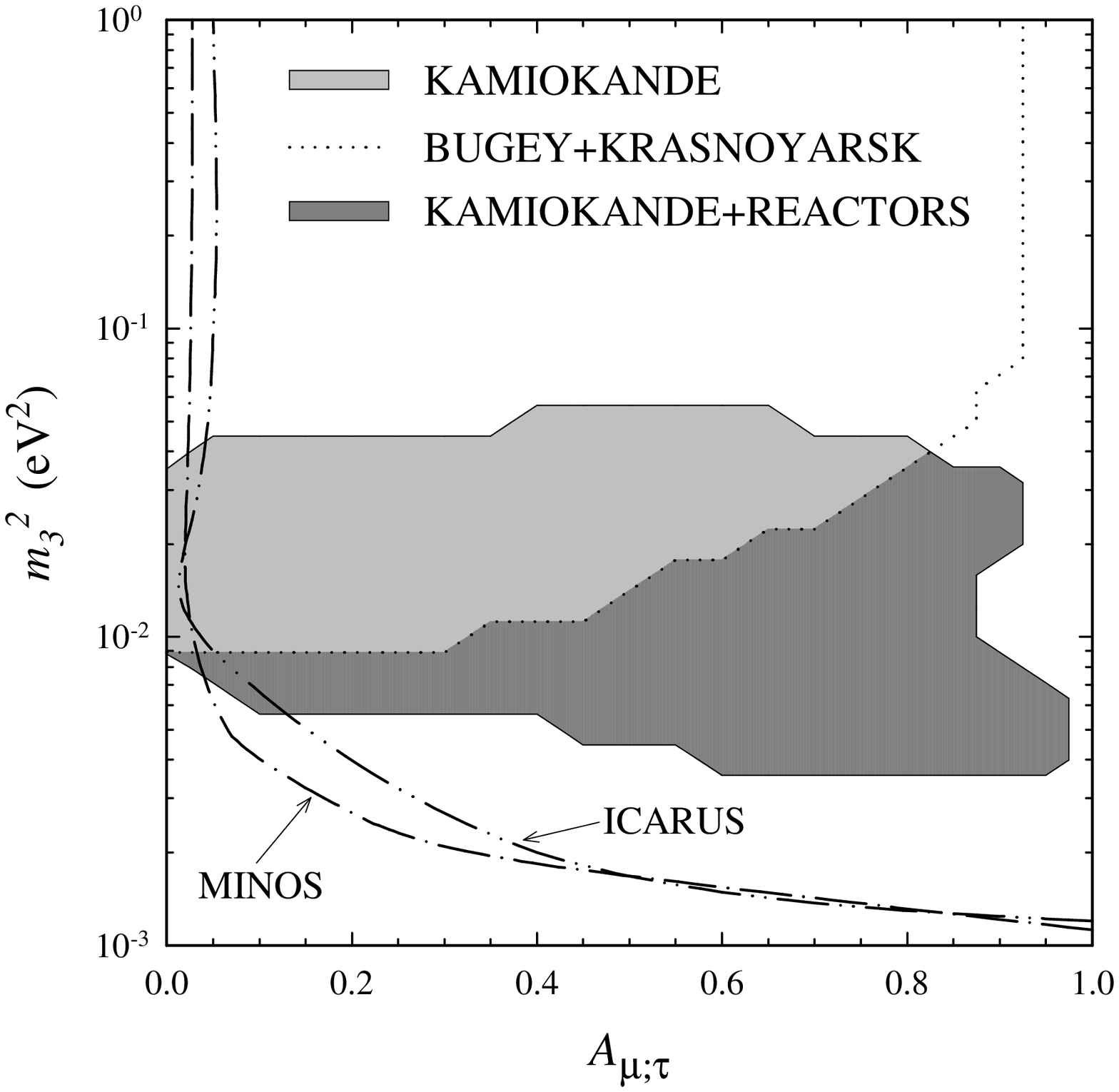}
\\[0.5cm]
Figure~\ref{fig7}
\end{center}
\end{figure}


\newpage

\begin{thebibliography}{99}

\bibitem{KAM88}
K.S. Hirata \textit{et al.},
Phys. Lett. B \textbf{205}, 416 (1988).

\bibitem{KAM92}
K.S. Hirata \textit{et al.},
Phys. Lett. B \textbf{280}, 146 (1992).

\bibitem{KAM94}
Y. Fukuda \textit{et al.},
Phys. Lett. B \textbf{335}, 237 (1994).

\bibitem{IMB}
D. Casper  \textit{et al.},
Phys. Rev. Lett. \textbf{66}, 2561 (1991);
R. Becker-Szendy \textit{et al.},
Phys. Rev. D \textbf{46}, 3720 (1992);
Nucl. Phys. B (Proc. Suppl.) \textbf{38}, 331 (1995).

\bibitem{Soudan}
T. Kafka,
Nucl. Phys. B (Proc. Suppl.) \textbf{35}, 427 (1994);
M. Goodman,
\textit{ibid.} \textbf{38}, 337 (1995);
W.W.M. Allison \textit{et al.},
Phys. Lett. B \textbf{391}, 491 (1997).

\bibitem{Pontecorvo57}
B. Pontecorvo,
J. Exptl. Theoret. Phys. \textbf{33}, 549 (1957)
[Sov. Phys. JETP \textbf{6}, 429 (1958)];
J. Exptl. Theoret. Phys. \textbf{34}, 247 (1958)
[Sov. Phys. JETP \textbf{7}, 172 (1958)].

\bibitem{BP78}
S.M. Bilenky and B. Pontecorvo,
Phys. Rep. \textbf{41}, 225 (1978).

\bibitem{BP87}
S.M. Bilenky and S.T. Petcov,
Rev. Mod. Phys. \textbf{59}, 671 (1987).

\bibitem{Mohapatra-Pal}
R.N. Mohapatra and P.B. Pal,
\textit{Massive Neutrinos in Physics and
Astrophysics},
World Scientific Lecture Notes in Physics, Vol.41
(World Scientific, Singapore, 1991).

\bibitem{CWKim}
C.W. Kim and A. Pevsner,
\textit{ Neutrinos in Physics and Astrophysics},
Contemporary Concepts in Physics, Vol.8
(Harwood Academic Press, Chur, Switzerland, 1993).

\bibitem{GR95}
G. Gelmini and E. Roulet,
Rept. Prog. Phys. \textbf{58}, 1207 (1995).

\bibitem{BGS}
S.M. Barr, T.K. Gaisser, P. Lipari and S. Tilav,
Phys. Lett. B \textbf{214}, 147 (1988);
G. Barr, T.K. Gaisser and T. Stanev,
Phys. Rev. D \textbf{39}, 3532 (1989);
W. Frati, T.K. Gaisser, A.K. Mann and T. Stanev,
Phys. Rev. D \textbf{48}, 1140 (1993);
T.K. Gaisser,
Nucl. Phys. B (Proc. Suppl.) \textbf{35}, 209 (1994).

\bibitem{Gaisser90}
T.K. Gaisser,
\textit{Cosmic Rays and Particle Physics}
(Cambridge University Press, 1990).

\bibitem{Naumov}
E.V. Bugaev and V.A. Naumov,
Yad. Fiz. \textbf{45}, 1380 (1987)
[Sov. J. Nucl. Phys. \textbf{45}, 857 (1987)];
Phys. Lett. B \textbf{232}, 391 (1989).

\bibitem{Honda}
M. Honda \textit{et al.},
Phys. Lett. B \textbf{248}, 193 (1990);
Phys. Rev. D \textbf{52}, 4985 (1995).

\bibitem{Perkins}
D.H. Perkins,
Astropart. Phys. \textbf{2}, 249 (1994).

\bibitem{SK-atm}
K. Martens,
Talk presented at the
\textit{International Europhysics Conference on High Energy Physics},
19-26 August 1997, Jerusalem, Israel
(http://\-www.cern.ch/\-hep97/\-abstract/\-tpa10.htm);
E. Kearns,
Talk presented at the Conference on
\textit{Solar Neutrinos: News About SNUs},
2--6 December 1997, Santa Barbara, California
(http://\-doug-pc.\-itp.\-ucsb.\-edu/\-online/\-snu/).

\bibitem{OR}
O.G. Ryazhskaya,
JETP Lett. \textbf{60}, 617 (1994 );
\textit{ibid.} \textbf{61}, 237 (1995).

\bibitem{KAM96}
Y. Fukuda \textit{et al.},
Phys. Lett. B \textbf{388}, 397 (1996).

\bibitem{Frejus}
C. Berger \textit{et al.},
Phys. Lett. B \textbf{227}, 489 (1989);
\textit{ibid.} \textbf{245}, 305 (1990);
K. Daum \textit{et al.},
Z. Phys. C \textbf{66}, 417 (1995).

\bibitem{NUSEX}
M. Aglietta \textit{et al.},
Europhys. Lett. \textbf{8}, 611 (1989);
\textit{ibid.} \textbf{15}, 559 (1991).

\bibitem{IMB97}
R. Clark \textit{et al.},
Phys. Rev. Lett. \textbf{79}, 345 (1997).

\bibitem{Fogli1}
G.L. Fogli, E. Lisi and D. Montanino,
Phys. Rev. D \textbf{49}, 3626 (1994);
Astrop. Phys. \textbf{4}, 177 (1995),
G.L. Fogli, E. Lisi, D. Montanino and G. Scioscia,
Phys. Rev. D \textbf{55}, 4385 (1997).

\bibitem{BGK}
S.M. Bilenky, C. Giunti and C.W. Kim,
Astrop. Phys. \textbf{4}, 241 (1996).

\bibitem{Kam-atm-up}
M. Mori \textit{et al.},
Phys. Lett. B \textbf{270}, 89 (1991).

\bibitem{IMB-up}
R. Becker-Szendy \textit{et al.},
Phys. Rev. Lett. \textbf{69}, 1010 (1992).

\bibitem{BAKSAN}
M.M. Boliev \textit{et al.},
\textit{Proc. of the
$3^{\mathrm{th}}$ International Workshop
on Neutrino Telescopes},
Venezia, March 1991.

\bibitem{MACRO}
MACRO Coll.,
S. Ahlen \textit{et al.},
Phys. Lett. B \textbf{357}, 481 (1995);
A. Surdo,
Talk presented at the
XVI$^{\mathrm{th}}$
\textit{International Workshop on
Weak Interactions and Neutrinos},
Capri, Italy, 22--28 June 1997;
MACRO Coll.,
M. Ambrosio \textit{et al.},
preprint INFN/AE-97/21.

\bibitem{NGS}
W. Frati \textit{et al.},
Phys. Rev. D \textbf{48}, 1140 (1993);
V. Agrawal \textit{et al.},
\textit{ibid.} \textbf{53}, 1314 (1996);
T.K. Gaisser \textit{et al.},
\textit{ibid.} \textbf{54}, 5578 (1996).

\bibitem{FLM97}
G.L. Fogli, E. Lisi and A. Marrone,
Phys. Rev. D \textbf{57}, 5893 (1998).

\bibitem{CHOOZ}
C. Bemporad,
Proc. of
\textit{Neutrino 96},
Helsinki, Finland, 13--19 June 1996,
edited by K. Enqvist, K. Huitu and J. Maalampi
(World Scientific, Singapore, 1997),
p.242;
http://\-duphy4.physics.drexel.edu/\-chooz\_pub/\-index.htmlx.

\bibitem{PaloVerde}
F. Boehm \textit{et al.},
\textit{The Palo Verde experiment},
1996
(http:\-//\-www.\-cco.\-caltech.\-edu/\~{}songhoon/\-Palo-Verde/\-Palo-Verde.\-html);
G. Gratta,
Proc. of
\textit{Neutrino 96},
Helsinki, Finland, 13--19 June 1996,
edited by K. Enqvist, K. Huitu and J. Maalampi
(World Scientific, Singapore, 1997),
p.248.

\bibitem{K2K}
Y. Suzuki,
Proc. of
\textit{Neutrino 96},
Helsinki, Finland, 13--19 June 1996,
edited by K. Enqvist, K. Huitu and J. Maalampi
(World Scientific, Singapore, 1997),
p.237;
http://\-pnahp.kek.jp/.

\bibitem{MINOS}
MINOS Coll.,
D. Ayres \textit{et al.},
NUMI-L-63, February 1995;
S.G. Wojcicki,
Proc. of
\textit{Neutrino 96},
Helsinki, Finland, 13--19 June 1996,
edited by K. Enqvist, K. Huitu and J. Maalampi
(World Scientific, Singapore, 1997),
p.231;
http://\-www.hep.anl.gov/\-NDK/\-HyperText/\-numi.html.

\bibitem{ICARUS}
ICARUS Coll.,
P. Cennini \textit{et al.},
LNGS-94/99-I,
May 1994;
http://\-www.aquila.infn.it/\-icarus/.

\bibitem{see-saw}
M. Gell-Mann, P. Ramond, and R. Slansky,
in \textit{Supergravity},
ed. F. van Nieuwenhuizen and D. Freedman
(North Holland, Amsterdam, 1979), p.315;
T. Yanagida,
\textit{Proc. of the
Workshop on Unified Theory and the Baryon Number of the Universe},
KEK, Japan, 1979;
S. Weinberg,
Phys. Rev. Lett. \textbf{43}, 1566 (1979).

\bibitem{Homestake}
B.T. Cleveland \textit{et al.},
Nucl. Phys. B (Proc. Suppl.) \textbf{38}, 47 (1995).

\bibitem{Kamiokande}
K.S. Hirata \textit{et al.},
Phys. Rev. Lett. \textbf{65}, 1297 (1990);
Phys. Rev. D \textbf{44}, 2241 (1991).

\bibitem{GALLEX}
GALLEX Coll.,
Phys. Lett. B \textbf{285}, 376 (1992);
\textit{ibid.} \textbf{314}, 445 (1993);
\textit{ibid.} \textbf{327}, 377 (1994);
\textit{ibid.} \textbf{342}, 440 (1995);
\textit{ibid.} \textbf{388}, 384 (1996).

\bibitem{SAGE}
J.N. Abdurashitov \textit{et al.},
Phys. Lett. B \textbf{328}, 234 (1994);
Phys. Rev. Lett. \textbf{77}, 4708 (1996).

\bibitem{SK-sun}
K. Inoue,
Talk presented at \textit{TAUP97},
September 7-11, 1997,
Laboratori Nazionali del Gran Sasso, Assergi (Italy);
R. Svoboda,
Talk presented at the Conference on
\textit{Solar Neutrinos: News About SNUs},
2--6 December 1997, Santa Barbara, California
(http://\-doug-pc.\-itp.\-ucsb.\-edu/\-online/\-snu/).

\bibitem{Bahcall}
J.N. Bahcall and R. Ulrich,
Rev. Mod. Phys. \textbf{60}, 297 (1988);
J.N. Bahcall,
\textit{Neutrino Physics and Astrophysics}
(Cambridge University Press, 1989);
J.N. Bahcall and M.H. Pinsonneault,
Rev. Mod. Phys. \textbf{64}, 885 (1992);
J.N. Bahcall and M.H. Pinsonneault,
\textit{ibid.} \textbf{67}, 781 (1995);

\bibitem{Saclay}
S. Turck-Chi\`eze, S. Cahen, M. Cass\'e and C. Doom,
Astrophys. J. \textbf{335}, 415 (1988);
S. Turck-Chi\`eze and I. Lopes,
\textit{ibid.} \textbf{408}, 347 (1993);
S. Turck-Chi\`eze \textit{et al.},
Phys. Rep. \textbf{230}, 57 (1993).

\bibitem{CDF}
V. Castellani, S. Degl'Innocenti and G. Fiorentini,
Astron. Astrophys. \textbf{271}, 601 (1993);
S. Degl'Innocenti,
Univ. of Ferrara preprint
INFN-FE-07-93;
V. Castellani \textit{et al.},
Phys. Rep. \textbf{281}, 309 (1997).

\bibitem{DS96}
A. Dar and G. Shaviv,
Nucl. Phys. B (Proc. Suppl.) \textbf{48}, 335 (1996).

\bibitem{phenomenological}
V. Castellani \textit{et al.},
Astron. Astrophys. \textbf{271}, 601 (1993);
S.A. Bludman \textit{et al.},
Phys. Rev. D \textbf{49}, 3622 (1994);
V. Berezinsky,
Comm. Nucl. Part. Phys. \textbf{21}, 249 (1994);
J.N. Bahcall,
Phys. Lett. B \textbf{338}, 276 (1994).

\bibitem{MSW}
S.P. Mikheyev and A.Yu. Smirnov,
Yad. Fiz. \textbf{42}, 1441 (1985)
[Sov. J. Nucl. Phys. \textbf{42}, 913 (1985)];
Il Nuovo Cimento C \textbf{9}, 17 (1986).

\bibitem{SOLMSW}
GALLEX Coll.,
P. Anselmann \textit{et al.},
Phys. Lett. B \textbf{285}, 390 (1992);
X. Shi, D.N. Schramm and J.N. Bahcall,
Phys. Rev. Lett. \textbf{69}, 717 (1992);
P.I. Krastev and S.T. Petcov,
Phys. Lett. B \textbf{299}, 99 (1993);
G.L. Fogli and E. Lisi,
Astropart. Phys. \textbf{2}, 91 (1994);
N. Hata and P.G. Langacker,
Phys. Rev. \textbf{50}, 632 (1994);
G. Fiorentini \textit{et al.},
\textit{ibid.} \textbf{49}, 6298 (1994);
L.M. Krauss, E. Gates and M. White,
\textit{ibid.} \textbf{51}, 2631 (1995).

\bibitem{SOLVAC}
V. Barger, R.J.N. Phillips, and K. Whisnant,
Phys. Rev. Lett. \textbf{69}, 3135 (1992);
P.I. Krastev and S.T. Petcov,
\textit{ibid.} \textbf{72}, 1960 (1994).

\bibitem{PDG}
R.M. Barnett \textit{et al.},
Phys. Rev. D \textbf{54}, 1 (1996).

\bibitem{Breviews}
R.G.H. Robertson and D.A. Knapp,
Ann. Rev. Nucl. Part. Sci. \textbf{38}, 185 (1988);
N.A. Jelley,
Proc. of the
\textit{$7^{\mathrm{th}}$ Int. Workshop
on Neutrino Telescopes},
Venezia, February 1996, p.131.

\bibitem{Wolfenstein78}
L. Wolfenstein,
Phys. Rev. D \textbf{17}, 2369 (1978);
\textit{ibid.} \textbf{20}, 2634 (1979).

\bibitem{matter}
E.D. Carlson,
Phys. Rev. D \textbf{34}, 1454 (1986);
P.I. Krastev and S.T. Petcov,
Phys. Lett. B \textbf{205}, 84 (1988);
G. Auriemma \textit{et al.},
\textit{ibid.} \textbf{37}, 665 (1988);
E. Akhmedov, P. Lipari and M. Lusignoli,
Phys. Lett. B \textbf{300}, 128 (1993).

\bibitem{KP89}
T.K. Kuo and J. Pantaleone,
Rev. Mod. Phys. \textbf{61}, 937 (1989).

\bibitem{LSND}
C. Athanassopoulos \textit{et al.},
Phys. Rev. Lett. \textbf{75}, 2650 (1995);
\textit{ibid.} \textbf{77}, 3082 (1996).

\bibitem{no3}
C.Y. Cardall and G.M. Fuller,
Phys. Rev. D \textbf{53}, 4421 (1996);
O. Yasuda and H. Minakata,
preprint TMUP-HEL-9604
(hep-ph/9602386);
S.M. Bilenky, C. Giunti and W. Grimus,
Eur. Phys. J. C \textbf{1}, 247 (1998);
N. Okada and O. Yasuda,
Int. J. Mod. Phys. A \textbf{12}, 3669 (1997).

\bibitem{KARMEN}
J. Kleinfeller,
Nucl. Phys. B (Proc. Suppl.) \textbf{48}, 207 (1996).

\bibitem{LSNDcheck}
L. Ludovici and P. Zucchelli,
preprint CERN-PPE/96-181;
Booster Neutrino Experiment
(BooNE),
http://\-nu1.lampf.lanl.gov/\-BooNE.

\bibitem{Boehm-Vannucci}
F. Boehm,
Nucl. Phys. B (Proc. Suppl.) \textbf{48}, 148 (1996);
F. Vannucci,
\textit{ibid.} \textbf{48}, 154 (1996).

\bibitem{Bugey}
B. Achkar \textit{et al.},
Nucl. Phys. B \textbf{434}, 503 (1995).

\bibitem{Krasnoyarsk}
G.S. Vidyakin \textit{et al.},
JETP Lett. \textbf{59}, 390 (1994).

\bibitem{KM73}
M. Kobayashi and T. Maskawa,
Prog. Theor. Phys. \textbf{49}, 652 (1973).

\bibitem{SV80}
J. Schechter and J.W.F. Valle,
Phys. Rev. D \textbf{21}, 309 (1980);
\textit{ibid.} \textbf{22}, 2227 (1980).

\bibitem{BHP80}
S.M. Bilenky, J. Hosek and S.T. Petcov,
Phys. Lett. B \textbf{94}, 495 (1980).

\bibitem{KMOS80}
I.Yu. Kobzarev \textit{et al.},
Yad. Fiz. \textbf{32}, 1590 (1980)
[Sov. J. Nucl. Phys. \textbf{32}, 823 (1980)].

\bibitem{Doi81}
M. Doi \textit{et al.},
Phys. Lett. B \textbf{102}, 323 (1981).

\bibitem{Langacker87}
P. Langacker \textit{et al.},
Nucl. Phys. B \textbf{282}, 589 (1987).

\bibitem{Cabibbo78}
N. Cabibbo,
Phys. Lett. B \textbf{72}, 333 (1978).

\bibitem{Stacey}
F.D. Stacey,
\textit{Physics of the Earth},
John Wiley \& Sons, Inc., 1977.

\bibitem{Anderson}
D.L. Anderson,
\textit{Theory of the Earth},
Blackwell Scientific Publications, 1989.

\bibitem{recent-sun}
N. Hata and P. Langacker,
Phys. Rev. D \textbf{56}, 6107 (1997);
G.L. Fogli, E. Lisi and D. Montanino,
preprint hep-ph/9709473.

\bibitem{future-sun}
Y. Suzuki (SuperKamiokande),
Proc. of the
\textit{Fourth International Solar Neutrino Conference},
Heidelberg, Germany, 8--11 April 1997,
edited by W. Hampel
(Max-Planck-Institut f\"ur Kernphysik, 1997),
p.163;
R. Meijer Drees (SNO),
\textit{ibid.}, p.210;
F. von Feilitzsch (Borexino),
\textit{ibid.}, p.192;
E. Bellotti (GNO),
\textit{ibid.}, p.173;
K. Lande (Homestake Iodine),
\textit{ibid.}, p.228;
C. Tao (HELLAZ),
\textit{ibid.}, p.238;
A.V. Kopylov (Lithium),
\textit{ibid.}, p.263;
Yu.G. Zdesenko (Xenon),
\textit{ibid.}, p.283;
P. Cennini \textit{et al.} (ICARUS),
LNGS-94/99-I,
May 1994;
T.J. Bowels (GaAs),
Proc. of
\textit{Neutrino 96},
Helsinki, Finland, 13--19 June 1996,
edited by K. Enqvist, K. Huitu and J. Maalampi
(World Scientific, Singapore, 1997),
p.83.

\bibitem{BFP92}
A. De Rujula \textit{et al.},
Nucl. Phys. B \textbf{168}, 54 (1980);
V. Barger and K. Whisnant,
Phys. Lett. B \textbf{209}, 365 (1988);
S.M. Bilenky, M. Fabbrichesi and S.T. Petcov,
\textit{ibid.} \textbf{276}, 223 (1992).

\bibitem{BBGK}
S.M. Bilenky \textit{et al.},
Phys. Lett. B \textbf{356}, 273 (1995);
Phys. Rev. D \textbf{54}, 1881 (1996).

\bibitem{Pantaleone94}
J. Pantaleone,
Phys. Rev. D \textbf{49}, R2152 (1994).

\bibitem{Pantaleone}
J. Pantaleone,
private communication
and Ref.\cite{Pantaleone94}.

\bibitem{Fogli2}
G.L. Fogli and E. Lisi,
Phys. Rev. D \textbf{52}, 2775 (1995).

\bibitem{tomography}
A. Nicolaidis,
Phys. Lett. B \textbf{200}, 553 (1988).

\bibitem{CDHS84}
F. Dydak \textit{et al.},
Phys. Lett. B \textbf{134}, 281 (1984).

\bibitem{Murnaghan62}
F.D. Murnaghan,
\textit{The unitary and rotation groups},
Spartan Books,
Washington,
D.C. 1962,
p.7.

\end{thebibliography}
\end{document}